\documentclass[journal]{IEEEtran}

\usepackage{amsmath}
\usepackage{amsfonts}
\usepackage{amssymb}
\usepackage{amstext}
\usepackage{amsthm}
\usepackage{gensymb}
\usepackage{cite}
\usepackage{graphicx}
\newtheorem{remark}{Remark}
\newtheorem{theorem}{Theorem}
\newtheorem{proposition}{Proposition}
\newtheorem{lemma}{Lemma}
\newtheorem{corollary}{Corollary}
\usepackage{algorithm}
\usepackage{algorithmic}
\usepackage{color}
\usepackage{makecell}
\usepackage{booktabs}

\title{Absolute User Fairness for a Downlink Multiuser MIMO-NOMA System}

\author{Mingyan~Gong 
        and~Bin~Lyu 
}

\begin{document}

\maketitle

\begin{abstract}
User fairness has been widely researched in various non-orthogonal multiple access (NOMA) systems. In this paper, max-min or min-max fairness is adopted for a downlink multiuser multiple-input multiple-output NOMA system, where one multi-antenna base station (BS) transmits to multiple NOMA clusters and each cluster is composed of two single-antenna users sharing a single beam on the same time-frequency resource block. Specifically, under perfect channel state information in the transmitter (CSIT), max-min fairness is adopted to maximize the minimum achievable rate among all users. Under partial CSIT, i.e., the BS only knows the distance and channel direction information of each user, min-max fairness is adopted to minimize the maximum outage probability among all users. Under either CSIT case, to avoid solving the coupled problem, zero-forcing beamforming is designed to obtain a power control problem and then a binary search or formula based algorithm is proposed. Theoretical analyses reveal that under either CSIT case, the problem can be guaranteed to have a unique solution and the proposed algorithm is computationally efficient due to only simple closed-form expressions involved. 

\end{abstract}


\IEEEpeerreviewmaketitle

\section{Introduction}

\IEEEPARstart{T}{he} exponential growth of connected devices and data-intensive applications has imposed unprecedented demands on wireless systems and driven the evolution toward the sixth generation \cite{Saad}. Apparently, these systems are required to deliver higher spectral efficiency, massive connectivity, lower latency, and equitable service of quality across heterogeneous user populations. Traditional orthogonal multiple access (OMA) schemes, which allocate exclusive time, frequency, or code resource blocks to individual users, have served as the dominant multiple access paradigm \cite{Tse,Goldsmith}. However, the strict orthogonality constraint fundamentally limits the number of servable users and constrains spectral utilization, particularly in some scenarios characterized by a high density of users with diverse channel conditions.

Non-orthogonal multiple access (NOMA) has emerged as a compelling alternative \cite{Islam,Dai,Maraqa}, which overcomes the orthogonality bottleneck by enabling multiple users to share the same time–frequency resource block through power-domain multiplexing. In a downlink NOMA system, the base station (BS) superimposes multiple users' signals with distinct power levels and users employ successive interference cancellation (SIC), which requires that users with good channel conditions sequentially decode and subtract the signals of users with poor channel conditions until their own information is extracted. Such architecture allows NOMA to achieve superior spectral efficiency and support massive connectivity. Notably, a widely acknowledged advantage of NOMA is its inherent capability to improve user fairness relative to OMA \cite{Saito} because it allocates a larger fraction of transmit power to users with poor channel conditions.


To data, several user fairness strategies have been widely researched in various NOMA systems, such as proportional fairness \cite{Chen,Liu,Choi} and $\alpha$-fairness \cite{Xu}. In particular, max–min fairness attracts more attention. As the name suggests, max-min fairness requires maximizing the minimum achievable rate or signal-to-interference-plus-noise ratio (SINR) among all users, thereby targeting the worst-case user experience. In  \cite{Timotheou}, max-min fairness is first applied to a downlink single-antenna NOMA system and a low-complexity
polynomial algorithm is developed to solve the power allocation problem. In \cite{Zhu}, max-min fairness is more complex because the problem involves the joint optimization of power allocation and channel assignment. To simplify this problem, the authors assume that each NOMA cluster contains only two users and thus design a simple bisection method. Additionally, max-min fairness is also applied in single-antenna NOMA systems with cooperative relaying \cite{Zhang,Mao}.

Compared to single-antenna NOMA systems, multiple-input multiple-output (MIMO) NOMA systems can further improve the spectral efficiency and massive connectivity of wireless systems by using the additional spatial dimension. Of course, integrating NOMA and MIMO presents novel challenges to max-min fairness, especially in the presence of downlink multiuser MIMO, in which the BS transmits independent data streams to multiple users on the same time–frequency resource block via space division multiple access \cite{Caire,Spencer,Peel,Shen,Yoo}. In fact, max-min fairness was applied to a downlink multiuser MIMO system in the early day \cite{schubert} and this requires us to solve a difficult coupled problem involving the joint optimization of power control and beamforming. Hence, max-min fairness in downlink multiuser MIMO-NOMA systems is also associated with this joint optimization. For example, the authors in \cite{Jiao} resort to alternating optimization to handle this joint optimization. However, existing algorithms for the joint optimization of power control and beamforming always have either or both of the following two flaws:
\begin{itemize}
    \item A good initial point must be given because the coupled problem is non-convex and generally has no closed-form solutions.
    \item The computational burden of the algorithm is high because each iteration requires eigen decomposition. 
\end{itemize}


In this paper, max-min or min-max fairness is adopted for a downlink multiuser MIMO-NOMA system, where one multi-antenna BS transmits to multiple NOMA clusters and each cluster is composed of two single-antenna users sharing a single beam on the same time-frequency resource block. Specifically, under perfect channel state information in the transmitter (CSIT), max-min fairness is adopted to maximize the minimum achievable rate among all users. Under partial CSIT, i.e., the BS only knows the distance and channel direction information (CDI) of each user, min-max fairness is adopted to minimize the maximum outage probability among all users. Under either CSIT case, to remedy both flaws mentioned and avoid solving the coupled problem, zero-forcing beamforming (ZFBF) is designed to obtain a power control problem and then a binary search or formula based algorithm is proposed. Theoretical analyses reveal that under either CSIT case, the problem can be guaranteed to have a unique solution and the proposed algorithm is computationally efficient due to only simple closed-form expressions involved. 

\emph{Notations}: $(\cdot)^T$, $(\cdot)^H$, $\vert\cdot\vert$, and $(\cdot)^{-1}$ mean the transpose, conjugate transpose, modulus, and inverse operators, respectively. $\Vert\mathbf{a}\Vert$ and $[\textbf{a}]_n$ stands for the Euclidean norm and $n$th element of vector $\mathbf{a}$, respectively. $(\cdot)^*$ denotes a/the global optimum of an optimization problem. $\mathrm{diag}\{a_1,\dots,a_N\}$ denotes the $N\times N$ diagonal matrix with diagonal elements $a_1,\dots,a_N$. Moreover, $\mathbf{1}=[1~\cdots~1]^T$ and $\mathbf{0}$ means an all-zero vector or matrix. $\mathcal{P}\{\cdot\}$ is the probability of an event and $\mathbf{I}_N$ is the identity matrix of order $N$.


\section{System Model}

\begin{figure}
\centerline{\includegraphics[width=22pc]{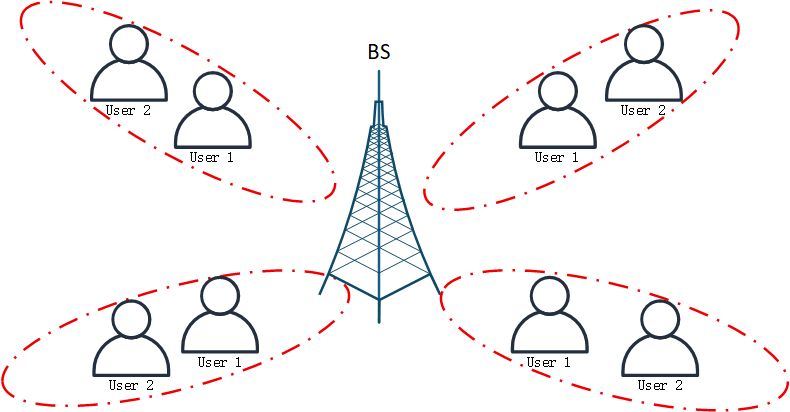}}
\caption{System model.}\label{system model}
\end{figure}

Let us consider a downlink multiuser MIMO-NOMA system, in which one BS equipped with $N$ antennas transmits to $2M~(M\le N)$ users each with one antenna using the same time–frequency resource block, as depicted in Fig. \ref{system model}. In order to implement multiuser MIMO and NOMA, these $2M$ users are grouped into $M$ clusters so that each cluster has two users sharing a single beam. In this context, beamforming is used to eliminate or mitigate inter-cluster interference while SIC is utilized to eliminate intra-cluster interference. We denote both users in the $n$th cluster by $\mathrm{UE}_{n,1}$ and $\mathrm{UE}_{n,2}$, respectively. Thus, the superimposed signal transmitted by the BS is
\begin{eqnarray}    \label{1}  
\mathbf{x}=\mathbf{W}\mathbf{s}=\sum_{m=1}^M\mathbf{w}_ms_m=\sum_{m=1}^M\mathbf{w}_m(s_{m,1}+s_{m,2})\in\mathbb{C}^{N},
\end{eqnarray}
where $\mathbf{W}=[\mathbf{w}_1~\cdots~\mathbf{w}_M]$ represents the precoding matrix, $\mathbf{w}_m$ stands for the unit-norm beamforming vector associated with the signal $s_m\sim\mathcal{CN}(0,p_m)$, $\mathbf{s}=[s_1~\cdots~s_M]^T$, and $s_{m,i}\sim\mathcal{CN}(0,p_{m,i})$ represents the information bearing signal for $\mathrm{UE}_{m,i}$. Here, all the $s_{m,i}$'s are assumed to be mutually uncorrelated and $p_m=p_{m,1}+p_{m,2}\ge0$ is the signal power of the $m$th cluster.

We assume that all wireless links exhibit statistically independent flat Rayleigh fading, and thus the signal received by $\mathrm{UE}_{n,i}$ can be expressed as
\begin{eqnarray}  \label{2}  
&&y_{n,i}=\mathbf{h}^H_{n,i}\mathbf{x}+v_{n,i}\nonumber\\
&&=\mathbf{h}^H_{n,i}\mathbf{w}_ns_{n,1}+\mathbf{h}^H_{n,i}\mathbf{w}_ns_{n,2}+\mathbf{h}^H_{n,i}\sum_{m\ne n}\mathbf{w}_ms_m+v_{n,i},\nonumber\\
&&n\in\mathcal{M}= \{1,2,\dots,M\},i\in\mathcal{I}=\{1,2\},
\end{eqnarray}
where $\mathbf{h}^H_{n,i}\sim\mathcal{CN}(\mathbf{0},\sigma^2_{n,i}\mathbf{I}_N)$ is the channel fading vector (i.e., CSIT) from the BS to $\mathrm{UE}_{n,i}$ and $v_{n,i}\sim\mathcal{CN}(0,1)$ is Gaussian noise. Moreover, $\sigma^2_{n,i}=d_{n,i}^{-\alpha}$ represents the large scale fading coefficient determined by the distance $d_{n,i}\ge1$ between the BS and $\mathrm{UE}_{n,i}$, and $\alpha$ is the path-loss coefficient. Importantly, let
\begin{eqnarray}
\mathbf{h}^H_{n,i}=\Vert\mathbf{h}_{n,i}\Vert\frac{\mathbf{h}^H_{n,i}}{\Vert\mathbf{h}_{n,i}\Vert}=\Gamma_{n,i}^{1/2}\mathbf{d}^H_{n,i}, n\in\mathcal{M},i\in\mathcal{I},
\end{eqnarray}
where $\Gamma_{n,i}^{1/2}=\Vert\mathbf{h}_{n,i}\Vert$ and $\mathbf{d}^H_{n,i}=\mathbf{h}^H_{n,i}/\Vert\mathbf{h}_{n,i}\Vert$ are the channel quality information (CQI) and CDI of $\mathrm{UE}_{n,i}$, respectively \cite{Yoo2,Gongone}. In this paper, each user only knows its own CSI but for the BS, we consider two cases about CSIT, namely,
\begin{itemize}
    \item perfect CSIT that the BS knows all the $\mathbf{h}^H_{n,i}$'s
    \item partial CSIT that the BS only knows the distance and CDI of each user without the CQI
\end{itemize}

\subsection{SIC Implementation}

SIC is utilized to eliminate intra-cluster interference, so we let the first user in each cluster use SIC. To this end, $\mathrm{UE}_{n,1}$ first treats its own signal $s_{n,1}$ as interference and decodes $\mathrm{UE}_{n,2}$'s signal $s_{n,2}$ by
\begin{eqnarray}  \label{3}  
y_{n,1}
=\overbrace{\mathbf{h}^H_{n,1}\mathbf{w}_ns_{n,1}}^{\mathrm{intra-cluster~interference}}+\overbrace{\mathbf{h}^H_{n,1}\mathbf{w}_ns_{n,2}}^{\mathrm{desired~signal}}\nonumber\\
+\underbrace{\mathbf{h}^H_{n,1}\sum_{m\ne n}\mathbf{w}_ms_m}_{\mathrm{inter-cluster~interference}}+v_{n,1}.
\end{eqnarray}
This requires $\mathrm{UE}_{n,2}$'s rate $R_{n,2}$ to satisfy
\begin{eqnarray}  \label{4}  
R_{n,2}\le\log_2\big(1+\mathrm{SINR}_{n,1}^2\big)~\mathrm{(bps/Hz)},
\end{eqnarray} 
where $C_{n,1}^2=\log_2\big(1+\mathrm{SINR}_{n,1}^2\big)$ and
\begin{eqnarray}
\mathrm{SINR}_{n,1}^2&=&\frac{p_{n,2}\vert\mathbf{h}^H_{n,1}\mathbf{w}_n\vert^2}{p_{n,1}\vert\mathbf{h}^H_{n,1}\mathbf{w}_n\vert^2+\sum_{m\ne n}p_m\vert\mathbf{h}^H_{n,1}\mathbf{w}_m\vert^2+1}\nonumber\\
&=&\frac{p_{n,2}}{p_{n,1}+\sum_{m\ne n}p_m\frac{\rho^2_{n,1,m}}{\rho^2_{n,1,n}}+\frac{1}{\Gamma_{n,1}\rho^2_{n,1,n}}},\nonumber
\end{eqnarray}
where $\Gamma_{n,i}>0$, $\rho^2_{n,i,m}=\vert\mathbf{h}^H_{n,i}\mathbf{w}_m\vert^2/\Gamma_{n,i}=\vert\mathbf{d}^H_{n,i}\mathbf{w}_m\vert^2$ measures the correlation between $\mathbf{h}_{n,i}$ and $\mathbf{w}_m$\footnote{In practice, each $\mathbf{h}_{n,i}\ne\mathbf{0}$ and each $\rho^2_{n,i,n}>0$ always hold and are assumed for analytical convenience in this paper.}.
After successfully decoding $s_{n,2}$, $\mathrm{UE}_{n,1}$ subtracts it from $y_{n,1}$ and obtains the residual signal
\begin{equation} \label{5} 
\tilde{y}_{n,1}=\mathbf{h}^H_{n,1}\mathbf{w}_ns_{n,1}+\mathbf{h}^H_{n,1}\sum_{m\ne n}\mathbf{w}_ms_m+v_{n,1}.
\end{equation}
This indicates that when $\mathrm{UE}_{n,1}$ decodes its own signal $s_{n,1}$, its rate $R_{n,1}$ satisfies
\begin{eqnarray}    \label{Cn1}
R_{n,1}\le\log_2\big(1+\mathrm{SINR}_{n,1}\big)~(\mathrm{bps/Hz}),
\end{eqnarray}
where $C_{n,1}=\log_2\big(1+\mathrm{SINR}_{n,1}\big)$ is $\mathrm{UE}_{n,1}$'s achievable rate and
\begin{eqnarray} \label{6}  %
\mathrm{SINR}_{n,1}&=&\frac{p_{n,1}\vert\mathbf{h}^H_{n,1}\mathbf{w}_n\vert^2}{\sum_{m\ne n}p_{m}\vert\mathbf{h}^H_{n,1}\mathbf{w}_m\vert^2+1}\nonumber\\
&=&\frac{p_{n,1}}{\sum_{m\ne n}p_m\frac{\rho^2_{n,1,m}}{\rho^2_{n,1,n}}+\frac{1}{\Gamma_{n,1}\rho^2_{n,1,n}}}\nonumber.
\end{eqnarray}

On the other hand, $\mathrm{UE}_{n,2}$ directly decodes its own signal $s_{n,2}$ without using SIC by
\begin{eqnarray} \label{7}  
y_{n,2}
=\overbrace{\mathbf{h}^H_{n,2}\mathbf{w}_ns_{n,1}}^{\mathrm{intra-cluster~interference}}+\overbrace{\mathbf{h}^H_{n,2}\mathbf{w}_ns_{n,2}}^{\mathrm{desired~signal}}\nonumber\\
+\underbrace{\mathbf{h}^H_{n,2}\sum_{m\ne n}\mathbf{w}_ms_m}_{\mathrm{inter-cluster~interference}}+v_{n,2}.
\end{eqnarray}
This requires that $\mathrm{UE}_{n,2}$'s rate $R_{n,2}$ also satisfies
\begin{equation} \label{8}  
R_{n,2}\le\log_2\big(1+\mathrm{SINR}_{n,2}^2\big)~\mathrm{(bps/Hz)},
\end{equation}
where $C_{n,2}^2=\log_2\big(1+\mathrm{SINR}_{n,2}^2\big)$ and
\begin{eqnarray}
\mathrm{SINR}_{n,2}^2&=&\frac{p_{n,2}\vert\mathbf{h}^H_{n,2}\mathbf{w}_n\vert^2}{p_{n,1}\vert\mathbf{h}^H_{n,2}\mathbf{w}_n\vert^2+\sum_{m\ne n}p_m\vert\mathbf{h}^H_{n,2}\mathbf{w}_m\vert^2+1},\nonumber\\
&=&\frac{p_{n,2}}{p_{n,1}+\sum_{m\ne n}p_m\frac{\rho^2_{n,2,m}}{\rho^2_{n,2,n}}+\frac{1}{\Gamma_{n,2}\rho^2_{n,2,n}}}.\nonumber
\end{eqnarray}

\subsection{Max-Min Fairness for Perfect CSIT}

In this subsection, we assume perfect CSIT such that the $C_{n,1}$'s and $C_{n,i}^2$'s are deterministic in the BS side and can be manipulated by user clustering, beamforming, and power control, prior to transmission. As a result, we use the max-min fairness strategy for users' achievable rates.

According to \eqref{4}, \eqref{Cn1}, and \eqref{8}, $\mathrm{UE}_{n,2}$'s achievable rate is
\begin{eqnarray}  \label{9}  
C_{n,2}=\min\big\{C_{n,1}^2,C_{n,2}^2\big\},n\in\mathcal{M}
\end{eqnarray}
which indicates that $C_{n,2}$ cannot achieve $C_{n,2}^2$ when $C_{n,1}^2<C_{n,2}^2$, possibly leading to unsuccessful SIC \cite{Sun,Yue}. To let $C_{n,2}$ achieve $C_{n,2}^2$, we give a necessary and sufficient condition in \textbf{Proposition \ref{p1}}.
\begin{proposition}  \label{p1}
    For the considered downlink multiuser MIMO-NOMA system with perfect CSIT, each $C_{n,2}$ can achieve $C_{n,2}^2$ if and only if
    \begin{eqnarray}     \label{10}  
    \rho^2_{n,2,n}\Big(\sum_{m\ne n}p_m\rho^2_{n,1,m}+\frac{1}{\Gamma_{n,1}}\Big)\nonumber~~~~~~~~~~~~\\
    \le\rho^2_{n,1,n}\Big(\sum_{m\ne n}p_m\rho^2_{n,2,m}+\frac{1}{\Gamma_{n,2}}\Big),\forall n\in\mathcal{M}.
    \end{eqnarray}
\end{proposition}

\begin{proof}
    According to \eqref{9}, we know that if $C_{n,2}=C_{n,2}^2$, we have $C_{n,1}^2\ge C_{n,2}^2\Leftrightarrow\mathrm{SINR}_{n,1}^2\ge\mathrm{SINR}_{n,2}^2$ and then \eqref{10} can be derived.
    The proof is completed.
\end{proof}

It can be seen from \textbf{Proposition \ref{p1}} that the inequality \eqref{10} involves the powers $p_m$'s (power control) and the correlations $\rho^2_{n,i,m}$'s (beamforming and user clustering). Unfortunately, it is well-known that the optimal user clustering for a NOMA system requires an exhaustive search among all users \cite{kimy2013non}, so we skip user clustering and analyze it in section \ref{sr.1}. When all users have been grouped and \eqref{10} is considered, the max-min fairness problem is designed as \cite{schubert,Jiao}
\begin{subequations} \label{maxmin1}
\begin{eqnarray}
&&\max_{\mathbf{P},\mathbf{W}}\min_{n\in\mathcal{M}}\big\{C_{n,1},C_{n,2}\big\}\\
&&~~~~\mathrm{s.t.}~~~p_{n,i}\ge0,n\in\mathcal{M},i\in\mathcal{I}\\
&&~~~~~~~~~~~\mathbf{1}^T\mathbf{P}\le P_{\mathrm{max}}\\
&&~~~~~~~~~~~\Vert\mathbf{w}_n\Vert=1,n\in\mathcal{M}\\
&&~~~~~~~~~~~\text{\eqref{10}}
\end{eqnarray}
\end{subequations}
where $\mathbf{P}=\big[p_{1,1}~p_{1,2}~\cdots~p_{M,1}~p_{M,2}\big]^T$ and $P_{\mathrm{max}}>0$ is the total transmission power available at the BS.


\subsection{Min-Max Fairness for Partial CSIT}

In this subsection, we assume partial CSIT that the BS does not know the CQI of each user, which makes the $C_{n,1}$'s and $C_{n,i}^2$'s random in the BS side. As a result, we use the min-max fairness strategy for users' outage probabilities \cite{Timotheou}.

Based on the outage probability definition in NOMA \cite{Ding,Gong2}, we analyze the statistical distribution of $\Gamma_{n,i}$ by
\begin{eqnarray}
    \Gamma_{n,i}=\sigma_{n,i}^2\Big\Vert\frac{\mathbf{h}_{n,i}}{\sigma_{n,i}}\Big\Vert^2=\sigma_{n,i}^2\xi_{n,i},\frac{\mathbf{h}_{n,i}}{\sigma_{n,i}}\sim\mathcal{CN}(\mathbf{0},\mathbf{I}_N).
\end{eqnarray}
Note that $2\xi_{n,i}$ is submit to the Chi-square distribution with $2N$ degrees of freedom, so all the $\xi_{n,i}$'s are independent and identically distributed and the probability density function of $\xi_{n,i}$ is
\begin{eqnarray}
    f_{\xi_{n,i}}(x)=\frac{1}{\Gamma(N)}x^{N-1}e^{-x},x>0,\forall n\in\mathcal{M},i\in\mathcal{I}
\end{eqnarray}
where $\Gamma(\cdot)$ is the gamma function, $\mathrm{SINR}_{n,i}^2$ and $\mathrm{SINR}_{n,1}$ can be regarded as random variables with respect to $\xi_{n,i}$:
\begin{subequations}
    \begin{eqnarray}
    \mathrm{SINR}_{n,i}^2
=\frac{p_{n,2}}{p_{n,1}+\sum_{m\ne n}p_m\frac{\rho^2_{n,i,m}}{\rho^2_{n,i,n}}+\frac{1}{\xi_{n,i}\sigma_{n,i}^2\rho^2_{n,iii,n}}}\in(0, u_{n,i}),\nonumber\\
        \mathrm{SINR}_{n,1}
=\frac{p_{n,1}}{\sum_{m\ne n}p_m\frac{\rho^2_{n,1,m}}{\rho^2_{n,1,n}}+\frac{1}{\xi_{n,1}\sigma_{n,1}^2\rho^2_{n,1,n}}}\in(0,u_n),~~~~~~~~\nonumber
    \end{eqnarray}
\end{subequations}
where $u_{n,i}=p_{n,2}/\big(p_{n,1}+\sum_{m\ne n}p_m\frac{\rho^2_{n,i,m}}{\rho^2_{n,i,n}}\big)$ and $u_n=p_{n,1}/\big({\sum_{m\ne n}p_m\frac{\rho^2_{n,1,m}}{\rho^2_{n,1,n}}}\big)$. Accordingly, the outage probabilities of $\mathrm{UE}_{n,1}$ and $\mathrm{UE}_{n,2}$ are calculated by
\begin{subequations}  \label{oup}
\begin{eqnarray}
    P_{n,1}^\mathrm{O}&=&1-\mathcal{P}\Big\{\mathrm{SINR}_{n,1}^2\ge\gamma\bigcap\mathrm{SINR}_{n,1}\ge \gamma\Big\},\\
    P_{n,2}^\mathrm{O}&=&1-\mathcal{P}\Big\{\mathrm{SINR}_{n,2}^2\ge\gamma\Big\},
\end{eqnarray}
\end{subequations}
where $\gamma\in\big(0,\min_{n\in\mathcal{M},i\in\mathcal{I}}\{u_{n,i},u_i\}\big)$ means the prespecified targeted SINR for all users. Likewise, when all users have been grouped, the min-max fairness problem is formulated as \cite{Timotheou}
\begin{subequations}  \label{minmax1}
\begin{eqnarray}
&&\min_{\mathbf{P},\mathbf{W}}\max_{n\in\mathcal{M}}\big\{P_{n,1}^\mathrm{O},P_{n,2}^\mathrm{O}\big\}\\
&&~~~~~\mathrm{s.t.}~~~p_{n,i}\ge0,n\in\mathcal{M},i\in\mathcal{I}\\
&&~~~~~~~~~~~~\mathbf{1}^T\mathbf{P}\le P_{\mathrm{max}}\\
&&~~~~~~~~~~~~\Vert\mathbf{w}_n\Vert=1,n\in\mathcal{M}.
\end{eqnarray}
\end{subequations}

\section{Max-Min Fairness for Perfect CSIT}

In this section, we aim to solve problem \eqref{maxmin1}. Unfortunately, this problem is very complicated due to the coupled nature \cite{schubert,Jiao}. For simplifying this problem, we first design the precoding matrix to obtain a power control problem.

\subsection{Precoding Matrix Design}

To avoid solving problems involving the joint optimization of power control and beamforming, in classic work on downlink multi-user MIMO systems, the major focus has been on cancelling co-channel interference by applying ZFBF \cite{Caire,Spencer,Peel,Shen,Yoo}. Thus, we also adopt ZFBF to first design $\mathbf{W}$. 

According to \eqref{10}, we find that when
\begin{itemize}
    \item $\rho^2_{n,1,m}=0\le\rho^2_{n,2,m},m\ne n$
    \item $\Gamma_{n,1}\ge\Gamma_{n,2}$
\end{itemize}
we have
\begin{eqnarray}  \label{11}  
    \sum_{m\ne n}\underbrace{p_m\rho^2_{n,1,m}}_{=0}+\frac{1}{\Gamma_{n,1}}\le
    \sum_{m\ne n}\underbrace{p_m\rho^2_{n,2,m}}_{\ge0}+\frac{1}{\Gamma_{n,2}},n\in\mathcal{M}.
\end{eqnarray}
To this end, when performing user clustering, we let $\Gamma_{n,1}\ge\Gamma_{n,2}$, i.e., $\mathrm{UE}_{n,1}$ and $\mathrm{UE}_{n,2}$ are the strong and weak users, respectively, which explains why we let the first user in each cluster use SIC. Then, we utilize the ZFBF based on the strong users' CSI or channel matrix\footnote{The ZFBF based on the strong users' channel matrix has been extensively adopted in downlink multiuser MIMO-NOMA systems \cite{kimy2013non,Hao}.}
\begin{eqnarray} \label{12}  
\mathbf{H}_1=
\begin{bmatrix}
    \mathbf{h}^H_{1,1}\\
    \vdots\\
    \mathbf{h}^H_{M,1}
\end{bmatrix}\in\mathbb{C}^{M\times N}
\end{eqnarray}
and design the precoding matrix as
\begin{equation}   \label{13}    
\mathbf{W}=\mathbf{H}_1^H(\mathbf{H}_1\mathbf{H}_1^H)^{-1}\cdot\mathrm{diag}\Big\{\frac{1}{\Vert\mathbf{f}_1\Vert},\dots,\frac{1}{\Vert\mathbf{f}_M\Vert}\Big\},
\end{equation}
where $\mathbf{f}_m$ is the $m$th column of $\mathbf{H}_1^H(\mathbf{H}_1\mathbf{H}_1^H)^{-1}$ and $\mathbf{w}_m=\mathbf{f}_m/\Vert\mathbf{f}_m\Vert$, and
\begin{eqnarray}  \label{14}  
\vert\mathbf{h}^H_{i,1}\mathbf{w}_j\vert^2=\Gamma_{i,1}\rho^2_{i,1,j}=\left\{
\begin{array}{ll}
0, & i\ne j, \\
1/\Vert\mathbf{f}_i\Vert^2 & i=j.
\end{array}
\right.
\end{eqnarray}

Utilizing \eqref{14}, we can obtain the inequality \eqref{11} and $C_{n,1}=\log_2(1+\Gamma_{n,1}\rho^2_{n,1,n}p_{n,1})$, and simplify \eqref{10} into
\begin{eqnarray}   \label{cons0.1}
\frac{\rho^2_{n,2,n}}{\Gamma_{n,1}}\le\rho^2_{n,1,n}\Big(\sum_{m\ne n}p_m\rho^2_{n,2,m}+\frac{1}{\Gamma_{n,2}}\Big),\forall n\in\mathcal{M}
\end{eqnarray}
or the compact matrix form
\begin{eqnarray}   \label{cons1}
    \mathbf{B}\mathbf{p}+\mathbf{b}_2\ge\mathbf{b}_1,
\end{eqnarray}
where $\mathbf{b}_i=\big[\frac{1}{\Gamma_{1,i}\rho^2_{1,i,1}}~\cdots~\frac{1}{\Gamma_{M,i}\rho^2_{M,i,M}}\big]^T>\mathbf{0},i\in\mathcal{I}$, $\mathbf{p}=[p_1~\cdots~p_M]^T$, and 
\begin{eqnarray}
    \mathbf{B}=\begin{bmatrix}
        0 & \frac{\rho^2_{1,2,2}}{\rho^2_{1,2,1}} & \cdots & \frac{\rho^2_{1,2,M}} {\rho^2_{1,2,1}} \\
        \frac{\rho^2_{2,2,1}}{\rho^2_{2,2,2}} & 0 & \cdots & \frac{\rho^2_{2,2,M}}{\rho^2_{2,2,2}} \\
        \vdots & \vdots & \ddots & \vdots \\
        \frac{\rho^2_{M,2,1}}{\rho^2_{M,2,M}} & \frac{\rho^2_{M,2,2}}{\rho^2_{M,2,M}} & \cdots & 0
    \end{bmatrix}\ge\mathbf{0}\in\mathbb{R}^{M\times M}.\nonumber
\end{eqnarray}
In order to guarantee the inequality \eqref{cons0.1}, we may additionally require
\begin{equation}  \label{15}  
\rho^2_{n,1,n}\ge\rho^2_{n,2,n},\forall n\in\mathcal{M}
\end{equation}
which can be achieved by user clustering. Due to limited users, however, there may not be a user clustering strategy for \eqref{15}. To address this issue, when $\rho^2_{n,1,n}<\rho^2_{n,2,n}$, we can actually increase some $p_m$'s to guarantee the inequality \eqref{cons0.1} or \eqref{cons1}, which is always feasible when $P_{\mathrm{max}}$ and the channel correlation between both users in each cluster are large.

After designing $\mathbf{W}$, we can simplify problem \eqref{maxmin1} into the power control problem:
\begin{subequations} \label{maxmin2}
\begin{eqnarray}
&&\max_{\mathbf{P}}\min_{n\in\mathcal{M}}\big\{C_{n,1},C_{n,2}\big\}\\
&&~~~~~\mathrm{s.t.}~~~\mathbf{P}\ge\mathbf{0},\mathbf{1}^T\mathbf{P}\le P_{\mathrm{max}}\\
&&~~~~~~~~~~~~\mathbf{B}\mathbf{p}+\mathbf{b}_2\ge\mathbf{b}_1.
\end{eqnarray}
\end{subequations}

\subsection{Power Control}


To begin with, we introduce the auxiliary variable $\beta_{n,i}$ as the power allocation factor of $\mathrm{UE}_{n,i}$, which satisfies $p_{n,i}=p_n\beta_{n,i}$ and $\beta_{n,1}+\beta_{n,2}=1$. Next, we use the relation of a data rate and its SINR to rewrite problem \eqref{maxmin2} as
\begin{subequations} \label{maxmin3}
\begin{eqnarray}
&&G(P_{\mathrm{max}})=\max_{\mathbf{p},\boldsymbol{\Theta}}\min_{n\in\mathcal{M}}\mathrm{SINR}^{\mathrm{min}}_n(\mathbf{p},\boldsymbol{\beta}_n)\\
&&~~~~\mathrm{s.t.}~~~\mathbf{p}\ge\mathbf{0},\mathbf{1}^T\mathbf{p}\le P_{\mathrm{max}}\\
&&~~~~~~~~~~~\boldsymbol{\beta}_n\ge\mathbf{0},\mathbf{1}^T\boldsymbol{\beta}_n=1,n\in\mathcal{M}\\
&&~~~~~~~~~~~\mathbf{B}\mathbf{p}+\mathbf{b}_2\ge\mathbf{b}_1
\end{eqnarray}
\end{subequations}
where $\boldsymbol{\beta}_n=[\beta_{n,1}~\beta_{n,2}]^T$ and $\boldsymbol{\Theta}=(\boldsymbol{\beta}_1,\dots,\boldsymbol{\beta}_M)$. Moreover, 
\begin{eqnarray}  \mathrm{SINR}^{\mathrm{min}}_n(\mathbf{p},\boldsymbol{\beta}_n)=\min\big\{\mathrm{SNR}_{n,1}(p_n,\boldsymbol{\beta}_{n}),\mathrm{SINR}_{n,2}(\mathbf{p},\boldsymbol{\beta}_n)\big\},\nonumber\\
    \mathrm{SNR}_{n,1}(p_n,\boldsymbol{\beta}_{n})=\Gamma_{n,1}\rho^2_{n,1,n}p_n\beta_{n,1},~~~~~~~~~~~~~~~~~~~~~~~~~~~~\nonumber\\
\mathrm{SINR}_{n,2}(\mathbf{p},\boldsymbol{\beta}_n)=\frac{p_{n}\beta_{n,2}}{p_{n}\beta_{n,1}+\sum_{m\ne n}p_m\frac{\rho^2_{n,2,m}}{\rho^2_{n,2,n}}+\frac{1}{\Gamma_{n,2}\rho^2_{n,2,n}}}.~~\nonumber
\end{eqnarray}

To shed some light on a global maximizer of problem \eqref{maxmin3}, we give the following lemmas.
\begin{lemma} \label{lem0}
If $(\mathbf{p}^*,\boldsymbol{\Theta}^*)$ is a global maximizer of problem \eqref{maxmin3}, we then have
\begin{eqnarray}   \label{lem0.1}
p_n^*\ne0,\beta_{n,i}^*\ne0,\forall n\in\mathcal{M},i\in\mathcal{I}.
\end{eqnarray}

\begin{proof}
    This proof is simple because when the constraint \eqref{lem0.1} does not hold, $G(P_{\mathrm{max}})=0$ and some user's achievable rate equals zero. The proof is completed.
\end{proof}

\end{lemma}
\begin{lemma} \label{lem1}
If $(\mathbf{p}^*,\boldsymbol{\Theta}^*)$ is a global maximizer of problem \eqref{maxmin3}, we then have $\mathbf{1}^T\mathbf{p}^*=P_{\mathrm{max}}$.
\begin{proof}
    We use the proof by contradiction. If $\mathbf{1}^T\mathbf{p}^*<P_{\mathrm{max}}$, we can set $\mathrm{SINR}^{\mathrm{min}}_n(\lambda\mathbf{p}^*,\boldsymbol{\beta}_n^*)$ with $\lambda\in\big[1,P_{\mathrm{max}}/(\mathbf{1}^T\mathbf{p}^*)\big]$. Then, we can derive that $\mathrm{SINR}^{\mathrm{min}}_n(\lambda\mathbf{p}^*,\boldsymbol{\beta}_n^*)$ increases as $\lambda$ increases, so there exists $\lambda_0\in(1,P_{\mathrm{max}}/(\mathbf{1}^T\mathbf{p}^*)\big]$ such that
    \begin{eqnarray}
\mathrm{SINR}^{\mathrm{min}}_n(\lambda_0\mathbf{p}^*,\boldsymbol{\beta}_n^*)>\mathrm{SINR}^{\mathrm{min}}_n(\mathbf{p}^*,\boldsymbol{\beta}_n^*),\forall n\in\mathcal{M}
    \end{eqnarray}
    which indicates $\min_{n\in\mathcal{M}}\mathrm{SINR}^{\mathrm{min}}_n(\lambda_0\mathbf{p}^*,\boldsymbol{\beta}_n^*)>G(P_{\mathrm{max}})$ and $(\mathbf{p}^*,\boldsymbol{\Theta}^*)$ is not a global maximizer of problem \eqref{maxmin3}. This contradicts the assumption and the proof is completed.
\end{proof}
\end{lemma}

\begin{lemma}  \label{lem2}
If $(\mathbf{p}^*,\boldsymbol{\Theta}^*)$ is a global maximizer of problem \eqref{maxmin3}, we then have
\begin{eqnarray}   \label{le1}
G(P_{\mathrm{max}})=\mathrm{SINR}^{\mathrm{min}}_n(\mathbf{p}^*,\boldsymbol{\beta}_n^*),\forall n\in\mathcal{M}.
\end{eqnarray}
\end{lemma}

\begin{proof}
    We use the proof by contradiction. If the equality \eqref{le1} does not hold, after sorting we can assume
    \begin{eqnarray}
G(P_{\mathrm{max}})=\mathrm{SINR}^{\mathrm{min}}_1(\mathbf{p}^*,\boldsymbol{\beta}_1^*)=\cdots=\mathrm{SINR}^{\mathrm{min}}_i(\mathbf{p}^*,\boldsymbol{\beta}_i^*)\nonumber\\<\mathrm{SINR}^{\mathrm{min}}_{i+1}(\mathbf{p}^*,\boldsymbol{\beta}_{i+1}^*)\le\cdots\le\mathrm{SINR}^{\mathrm{min}}_M(\mathbf{p}^*,\boldsymbol{\beta}_M^*).
    \end{eqnarray}
    Then, we set $\sum_{m=1}^ip_m^*=P_f>0$, $\sum_{m=i+1}^Mp_m^*=P_s>0$, and $\mu P_f+\eta P_s=\sum_{m=1}^Mp_m^*$, where $\mu\in\big[1,(\sum_{m=1}^Mp_m^*)/P_f\big)$ and $\eta=(\sum_{m=1}^Mp_m^*-\mu P_f)/P_s\in(0,1]$ decreases as $\mu$ increases. On this foundation, we find that as $\mu$ increases, 
\begin{eqnarray}
  f_n(\mu)=\sum_{m=1,m\ne n}^i\frac{p_m^*\rho^2_{n,2,m}}{\rho^2_{n,2,n}}+\sum_{m=i+1}^M\frac{\eta}{\mu} \frac{p_m^*\rho^2_{n,2,m}}{\rho^2_{n,2,n}},n\le i 
\end{eqnarray}
decreases but
\begin{eqnarray}
    f_n(\mu)=\sum_{m=1}^i\frac{\mu}{\eta} \frac{p_m^*\rho^2_{n,2,m}}{\rho^2_{n,2,n}}+\sum_{m=i+1,m\ne n}^M \frac{p_m^*\rho^2_{n,2,m}}{\rho^2_{n,2,n}},n>i 
\end{eqnarray}
increases. Next, let $\mathbf{p}(\mu)=[\mu p_1^*~\cdots~\mu p_i^*~\eta p_{i+1}^*~\cdots~\eta p_M^*]^T$ and as $\mu$ increases,
    \begin{eqnarray}
\mathrm{SINR}^{\mathrm{min}}_n\big(\mathbf{p}(\mu),\boldsymbol{\beta}_n^*\big)=\min\Big\{\Gamma_{n,1}\rho^2_{n,1,n}(\mu p_n^*)\beta_{n,1}^*,\nonumber\\
        \frac{(\mu p_n^*)\beta_{n,2}^*}{(\mu p_n^*)\beta_{n,1}^*+\mu f_n(\mu)+\frac{1}{\Gamma_{n,2}\rho^2_{n,2,n}}}\Big\}\nonumber\\
        =\min\Big\{\Gamma_{n,1}\rho^2_{n,1,n}(\mu p_n^*)\beta_{n,1}^*,~~~~~~~~~~~~~~~~~~~\nonumber\\
        \frac{p_n^*\beta_{n,2}^*}{p_n^*\beta_{n,1}^*+f_n(\mu)+\frac{1}{\mu\Gamma_{n,2}\rho^2_{n,2,n}}}\Big\},n\le i
    \end{eqnarray}
    increases but
    \begin{eqnarray}
\mathrm{SINR}^{\mathrm{min}}_n\big(\mathbf{p}(\mu),\boldsymbol{\beta}_n^*\big)=\min\Big\{\Gamma_{n,1}\rho^2_{n,1,n}(\eta p_n^*)\beta_{n,1}^*,\nonumber\\
        \frac{(\eta p_n^*)\beta_{n,2}^*}{(\eta p_n^*)\beta_{n,1}^*+\eta f_n(\mu) +\frac{1}{\Gamma_{n,2}\rho^2_{n,2,n}}}\Big\}\nonumber\\
        =\min\Big\{\Gamma_{n,1}\rho^2_{n,1,n}(\eta p_n^*)\beta_{n,1}^*,~~~~~~~~~~~~~~~~~~\nonumber\\
        \frac{p_n^*\beta_{n,2}^*}{p_n^*\beta_{n,1}^*+f_n(\mu) +\frac{1}{\eta\Gamma_{n,2}\rho^2_{n,2,n}}}\Big\},n>i
    \end{eqnarray}
    decreases. Hence, there must exist $\mu_0\in\big(1,(\sum_{m=1}^Mp_m^*)/P_f\big)$ and $\eta_0=(\sum_{m=1}^Mp_m^*-\mu_0 P_f)/P_s$ such that
    \begin{eqnarray}
&&G(P_{\mathrm{max}})=\mathrm{SINR}^{\mathrm{min}}_l\big(\mathbf{p}(1),\boldsymbol{\beta}_l^*\big)<\mathrm{SINR}^{\mathrm{min}}_l\big(\mathbf{p}(\mu_0),\boldsymbol{
\beta}_l^*\big)\nonumber\\&&
\le\mathrm{SINR}^{\mathrm{min}}_k\big(\mathbf{p}(\mu_0),\boldsymbol{\beta}_k^*\big),\forall l\le i,k>i
    \end{eqnarray}
    which means $\min_{n\in\mathcal{M}}\mathrm{SINR}^{\mathrm{min}}_n\big(\mathbf{p}(\mu_0),\boldsymbol{\beta}_n^*\big)>G(P_{\mathrm{max}})$ and $(\mathbf{p}^*,\boldsymbol{\Theta}^*)$ is not a global maximizer of problem \eqref{maxmin3}. This contradicts the assumption and the proof is completed.
\end{proof}

Utilizing \textbf{Lemma \ref{lem0}}, \textbf{Lemma \ref{lem1}}, and \textbf{Lemma \ref{lem2}}, we simplify problem \eqref{maxmin3} into
\begin{subequations} \label{maxmin4}
\begin{eqnarray}
&&G(P_{\mathrm{max}})=\max_{\mathbf{p},\boldsymbol{\Theta},t}t\\
&&~~~~~\mathrm{s.t.}~~~\mathrm{SINR}^{\mathrm{min}}_n(\mathbf{p},\boldsymbol{\beta}_n)=t,n\in\mathcal{M}\\
&&~~~~~~~~~~~~\mathbf{p}>\mathbf{0},\mathbf{1}^T\mathbf{p}= P_{\mathrm{max}}\\
&&~~~~~~~~~~~~\boldsymbol{\beta}_n>\mathbf{0},\mathbf{1}^T\boldsymbol{\beta}_n=1,n\in\mathcal{M}\\
&&~~~~~~~~~~~~\mathbf{B}\mathbf{p}+\mathbf{b}_2\ge\mathbf{b}_1.\label{maxmin4.5}
\end{eqnarray}
\end{subequations}

In order to further simplify problem \eqref{maxmin4}, we provide the following lemma.
\begin{lemma}  \label{lem3}
    If $(\mathbf{p}^*,\boldsymbol{\Theta}^*,t^*)$ is a global maximizer of problem \eqref{maxmin4}, we then have
    \begin{eqnarray}  \label{le3.1}
        t^*=\mathrm{SNR}_{n,1}(p_n^*,\boldsymbol{\beta}_n^*)=\mathrm{SINR}_{n,2}(\mathbf{p}^*,\boldsymbol{\beta}_n^*),\forall n\in\mathcal{M}.
    \end{eqnarray}
\end{lemma}

\begin{proof}
    We still use the proof by contradiction. If the equality \eqref{le3.1} does not hold, there must exist $k\in\mathcal{M}$ such that either $\mathrm{SNR}_{k,1}(p_k^*,\boldsymbol{\beta}_k^*)>\mathrm{SINR}_{k,2}(\mathbf{p}^*,\boldsymbol{\beta}_k^*)$ or $\mathrm{SNR}_{k,1}(p_k^*,\boldsymbol{\beta}_k^*)<\mathrm{SINR}_{k,2}(\mathbf{p}^*,\boldsymbol{\beta}_k^*)$. Then, we note that as $\beta_{k,1}$ increases,
    \begin{eqnarray}   \label{lem3.2}
        \mathrm{SNR}_{k,1}(p_k^*,\boldsymbol{\beta}_k)=\Gamma_{k,1}\rho^2_{k,1,k}p_k^*\beta_{k,1}
    \end{eqnarray}
    increases but
        \begin{eqnarray}   \label{lem3.3}
        \mathrm{SINR}_{k,2}(\mathbf{p}^*,\boldsymbol{\beta}_k)=~~~~~~~~~~~~~~~~~~~~\nonumber\\
        \frac{p_k^*(1-\beta_{k,1})}{p_k^*\beta_{k,1}+\sum_{m\ne k}p_m^*\frac{\rho^2_{k,2,m}}{\rho^2_{k,2,k}}+\frac{1}{\Gamma_{k,2}\rho^2_{k,2,k}}}
    \end{eqnarray}
    decreases. Moreover, \textbf{Lemma \ref{lem0}} indicates $\beta_{k,1}^*\in(0,1)$. Hence, when $\mathrm{SNR}_{k,1}(p_k^*,\boldsymbol{\beta}_k^*)>\mathrm{SINR}_{k,2}(\mathbf{p}^*,\boldsymbol{\beta}_k^*)$, there exists $\boldsymbol{\beta}_k^0=[\beta_{k,1}^0~\beta_{k,2}^0]$ with $\beta_{k,1}^0\in(0,\beta_{k,1}^*)$ such that
    \begin{eqnarray}
        \mathrm{SNR}_{k,1}(p_k^*,\boldsymbol{\beta}_k^*)>\mathrm{SNR}_{k,1}(p_k^*,\boldsymbol{\beta}_k^0)\nonumber\\
        \ge\underbrace{\mathrm{SINR}_{k,2}(\mathbf{p}^*,\boldsymbol{\beta}_k^0)}_{=\mathrm{SINR}^{\mathrm{min}}_k(\mathbf{p}^*,\boldsymbol{\beta}_k^0)}>\underbrace{\mathrm{SINR}_{k,2}(\mathbf{p}^*,\boldsymbol{\beta}_k^*)}_{=\mathrm{SINR}^{\mathrm{min}}_k(\mathbf{p}^*,\boldsymbol{\beta}_k^*)},
    \end{eqnarray}
    which indicates that $\mathrm{SINR}^{\mathrm{min}}_k(\mathbf{p}^*,\boldsymbol{\beta}_k^0)>\mathrm{SINR}^{\mathrm{min}}_k(\mathbf{p}^*,\boldsymbol{\beta}_k^*)=t^*=G(P_{\mathrm{max}})$ and $(\mathbf{p}^*,\boldsymbol{\Theta}^*,t^*)$ is not a global maximizer of problem \eqref{maxmin4}. This contracts the assumption. 
    
    When $\mathrm{SNR}_{k,1}(p_k^*,\boldsymbol{\beta}_k^*)<\mathrm{SINR}_{k,2}(\mathbf{p}^*,\boldsymbol{\beta}_k^*)$, we can obtain the same contradiction. The proof is completed.
\end{proof}

\textbf{Lemma \ref{lem2}} and \textbf{Lemma \ref{lem3}} verify that max-min fairness always causes absolute user fairness \cite{Timotheou,Zhu}, i.e.,  all users' SINRs are the same $t^*$. Based on \textbf{Lemma \ref{lem3}}, we can obtain an important result given in \textbf{Theorem \ref{theo1}}.

\begin{theorem}  \label{theo1}
    If $(\mathbf{p}^*,\boldsymbol{\Theta}^*,t^*)$ is a global maximizer of problem \eqref{maxmin4}, we have
    \begin{subequations}
    \begin{eqnarray}  \label{the01.1}
        \beta_{n,1}^*&=&\frac{-A_n(\mathbf{p}^*)+\sqrt{A_n^2(\mathbf{p}^*)+\frac{4p_n^*}{\Gamma_{n,1}\rho^2_{n,1,n}}}}{2p_n^*},\\
        p_{n,1}^*&=&\beta_{n,1}^*p_n^*\nonumber\\
        &=&\frac{-A_n(\mathbf{p}^*)+\sqrt{A_n^2(\mathbf{p}^*)+\frac{4p_n^*}{\Gamma_{n,1}\rho^2_{n,1,n}}}}{2},
    \end{eqnarray}
    \end{subequations}
    where $\beta_{n,1}^*>0$, $p_{n,1}^*>0$, $n\in\mathcal{M}$, and
    \begin{eqnarray}
        A_n(\mathbf{p})=\sum_{m\ne n}p_m\frac{\rho^2_{n,2,m}}{\rho^2_{n,2,n}}+\frac{1}{\Gamma_{n,1}\rho^2_{n,1,n}}+\frac{1}{\Gamma_{n,2}\rho^2_{n,2,n}}.\nonumber
    \end{eqnarray}
\end{theorem}

\begin{proof}
    Let $\mathrm{SNR}_{k,1}(p_k^*,\boldsymbol{\beta}_k)$ in \eqref{lem3.2} equal $\mathrm{SNIR}_{k,2}(\mathbf{p}^*,\boldsymbol{\beta}_k)$ in \eqref{lem3.3} and obtain a quadratic equation with respect to $\beta_{k,1}$. Solving this equation obtains \eqref{the01.1} and the proof is completed.
\end{proof}

From \textbf{Theorem \ref{theo1}}, we can obtain the following corollary.

\begin{corollary} \label{coro1}
    If $(\mathbf{p}^*,\boldsymbol{\Theta}^*,t^*)$ is a global maximizer of problem \eqref{maxmin4}, we have $\beta_{n,1}^*<\frac{1}{2},\forall n\in\mathcal{M}$.
\end{corollary}

\begin{proof}
    Since $(\mathbf{p}^*,\boldsymbol{\Theta}^*,t^*)$ is a global maximizer of problem \eqref{maxmin4}, we utilize the constraint \eqref{cons1} or \eqref{maxmin4.5} to construct
    \begin{eqnarray}
        &&\sum_{m\ne n}p_m^*\frac{\rho^2_{n,2,m}}{\rho^2_{n,2,n}}+\frac{1}{\Gamma_{n,2}\rho^2_{n,2,n}}\ge\frac{1}{\Gamma_{n,1}\rho^2_{n,1,n}}\nonumber\\
        &\Rightarrow& A_n(\mathbf{p}^*)\ge\frac{2}{\Gamma_{n,1}\rho^2_{n,1,n}}
        \Rightarrow 2p_n^*A_n(\mathbf{p}^*)\ge\frac{4p_n^*}{\Gamma_{n,1}\rho^2_{n,1,n}}\nonumber\\
        &\Rightarrow& [p_n^*]^2+2p_n^*A_n(\mathbf{p}^*)>\frac{4p_n^*}{\Gamma_{n,1}\rho^2_{n,1,n}},\forall n\in\mathcal{M}.
    \end{eqnarray}
    Next, we utilize \eqref{the01.1} to derive
    \begin{eqnarray}
        \beta_{n,1}^*&<&\frac{-A_n(\mathbf{p}^*)+\sqrt{A_n^2(\mathbf{p}^*)+2p_n^*A_n(\mathbf{p}^*)+[p_n^*]^2}}{2p_n^*}\nonumber\\
        &=&\frac{1}{2},\forall n\in\mathcal{M}.
    \end{eqnarray}
    The proof is completed.
\end{proof}

\begin{remark}
In the considered downlink multiuser MIMO-NOMA system, SIC is performed according to the CQI order in each cluster, which is guaranteed by imposing the implicit power constraint $p_{n,1}<p_{n,2}$ or $\beta_{n,1}<\frac{1}{2},\forall n\in\mathcal{M}$. From the SIC perspective, a power difference between both users in each cluster is necessary \cite{Islam,Dai,Maraqa} and impotently the max-min fairness strategy can guarantee this implicit constraint.
\end{remark}

Utilizing \textbf{Theorem \ref{theo1}}, we simplify problem \eqref{maxmin4} into
\begin{subequations} \label{maxmin5}
\begin{eqnarray}
&&G(P_{\mathrm{max}})=\max_{\mathbf{p},t}t\\
&&~~~~~\mathrm{s.t.}~~~\mathrm{SINR}^{\mathrm{min}}_n(\mathbf{p})=t,n\in\mathcal{M}\label{maxmin5.2}\\
&&~~~~~~~~~~~~\mathbf{p}>\mathbf{0},\mathbf{1}^T\mathbf{p}= P_{\mathrm{max}}\\
&&~~~~~~~~~~~~\mathbf{B}\mathbf{p}+\mathbf{b}_2\ge\mathbf{b}_1\label{maxmin5.4}
\end{eqnarray}
\end{subequations}
where the feasible region of $t$, the same SINR for all users, is obviously a subset of $(0, \infty)$ and
\begin{eqnarray}
\mathrm{SINR}^{\mathrm{min}}_n(\mathbf{p})=\Gamma_{n,1}\rho^2_{n,1,n}\frac{-A_n(\mathbf{p})+\sqrt{A_n^2(\mathbf{p})+\frac{4p_n}{\Gamma_{n,1}\rho^2_{n,1,n}}}}{2}.\nonumber
\end{eqnarray}

\subsection{Iterative Solution}

To solve problem \eqref{maxmin5}, we first focus more on the equality constraint \eqref{maxmin5.2} and transform it into
\begin{eqnarray}
    \frac{p_n}{t}-\sum_{m\ne n}p_m\frac{\rho^2_{n,2,m}}{\rho^2_{n,2,n}}=\frac{t+1}{\Gamma_{n,1}\rho^2_{n,1,n}}+\frac{1}{\Gamma_{n,2}\rho^2_{n,2,n}},n\in\mathcal{M}
\end{eqnarray}
and the compact matrix form
\begin{eqnarray}  \label{SLs}
    \mathbf{A}(t)\mathbf{p}=(t+1)\mathbf{b}_1+\mathbf{b}_2,
\end{eqnarray}
where $\mathbf{A}(t)=\mathbf{I}_M/t-\mathbf{B}$. 
Next, we depend on the following proposition to analyze the invertibility of $\mathbf{A}(t)$.
\begin{proposition}   \label{pro2}
    $\mathbf{A}(t)$ is invertible if and only if $t\ne1/\lambda_i,i=1,\dots,L$, where $1\le L<M$, the $\lambda_i$'s are the eigenvalues of $\mathbf{B}$, and $\lambda_1=\cdots=\lambda_K>\cdots\ge\lambda_L>0\ge\lambda_{L+1}\ge\cdots\ge\lambda_M$.
\end{proposition}
\begin{proof}
    The eigen decomposition of $\mathbf{B}$ can be written as $\mathbf{B}=\mathbf{Q}\mathrm{diag}\{\lambda_1,\dots,\lambda_M\}\mathbf{Q}^{-1}$, where $\sum_{i=1}^M\lambda_i=0$ because all the diagonal elements of $\mathbf{B}$ equal zero. Accordingly, the $\lambda_i$'s can be sorted as $\lambda_1\ge\cdots\ge\lambda_L>0\ge\lambda_{L+1}\ge\cdots\ge\lambda_M$ and the eigenvalues of $\mathbf{A}(t)$ are the $(\frac{1}{t}-\lambda_i)$'s. Then, the invertible $\mathbf{A}(t)$ indicates that all its eigenvalues do not equal zero, i.e., $\frac{1}{t}-\lambda_i\ne0,i\le L$ or $t\ne\frac{1}{\lambda_i}>0,i\le L$. The proof is completed.
\end{proof}
\textbf{Proposition \ref{pro2}} indicates that only $L$ values of $t>0$ cause $\mathbf{A}(t)$ being non-invertible, so we consider $\mathbf{A}(t)$ being invertible. Given $t>0$ and the invertible $\mathbf{A}(t)$, solving the system of linear equations \eqref{SLs} obtains
\begin{eqnarray}  \label{pFor}
    \mathbf{p}(t)&=&\mathbf{A}^{-1}(t)\big[(t+1)\mathbf{b}_1+\mathbf{b}_2\big]\nonumber\\
    &=&(t+1)\sum_{i=1}^M\frac{\mathbf{q}_i\mathbf{g}_i^T\mathbf{b}_1}{\frac{1}{t}-\lambda_i}+\sum_{i=1}^M\frac{\mathbf{q}_i\mathbf{g}_i^T\mathbf{b}_2}{\frac{1}{t}-\lambda_i},
\end{eqnarray}
where $\mathbf{A}^{-1}(t)=\mathbf{Q}\mathrm{diag}\{(\frac{1}{t}-\lambda_1)^{-1},\dots,(\frac{1}{t}-\lambda_M)^{-1}\}\mathbf{Q}^{-1}$, $\mathbf{Q}=[\mathbf{q}_1~\cdots~\mathbf{q}_M]$, and $\mathbf{Q}^{-1}=[\mathbf{g}_1~\cdots~\mathbf{g}_M]^T$.
To proceed, we give a necessary and sufficient condition for $\mathbf{p}(t)>\mathbf{0}$ in \eqref{pFor} by the following lemma \cite[p. 670]{Meyer} and theorem.
\begin{lemma}  \label{lem4}
    For the non-negative real matrix $\mathbf{B}\ge\mathbf{0}$, its spectral radius $\rho(\mathbf{B})=\lambda_1$.
\end{lemma}
    
\begin{theorem}  \label{theo2}
    $\mathbf{p}(t)>\mathbf{0}$ in \eqref{pFor} if and only if $\rho(\mathbf{B})<\frac{1}{t}$ or $0<t<\frac{1}{\lambda_1}$\footnote{In practice, $\rho(\mathbf{B})=\lambda_1>0$ always holds and is assumed in this paper.}.
\end{theorem}

\begin{proof}
    We first verify the sufficiency. When $\rho(\mathbf{B})<\frac{1}{t}$ or $0<t<\frac{1}{\lambda_1}$ based on \textbf{Lemma \ref{lem4}}, we have $\rho(t\mathbf{B})<1$,
    \begin{eqnarray}  \label{theo2.1}
        \mathbf{A}^{-1}(t)=t(\mathbf{I}_M-t\mathbf{B})^{-1}=t\sum_{i=0}^\infty(t\mathbf{B})^i,
        \end{eqnarray}
        and
        \begin{eqnarray}  \label{theo2.2}
        \mathbf{p}(t)=\underbrace{t\big[(t+1)\mathbf{b}_1+\mathbf{b}_2\big]}_{>\mathbf{0}}+\sum_{i=1}^\infty \underbrace{t^{i+1}\mathbf{B}^i\big[(t+1)\mathbf{b}_1+\mathbf{b}_2}_{\ge\mathbf{0}}\big]
    \end{eqnarray}
    which shows $\mathbf{p}(t)>\mathbf{0}$ due to $t>0$, $\mathbf{B}\ge\mathbf{0}$, and $\mathbf{b}_i>\mathbf{0}$.
    
    On the other hand, the necessity can be verified by the proof of \textbf{Theorem 2} in \cite{Boche}. The proof is completed.
\end{proof}

    \textbf{Theorem \ref{theo2}} indicates $t\in(0,1/\lambda_1)$, $G(P_{\mathrm{max}})\in(0,1/\lambda_1)$, and the invertible $\mathbf{A}(t)$ in \eqref{SLs}. That is to say, after user clustering and beamforming, $\mathbf{B}$ is determined and the objective function of problem \eqref{maxmin5} has an upper bound, namely, $t<\frac{1}{\lambda_1}\Rightarrow G(P_{\mathrm{max}})<\frac{1}{\lambda_1}$.
Moreover, \textbf{Theorem \ref{theo2}} directly causes the following corollary and proposition.
\begin{corollary}  \label{coro2}
    When $0<t<1/\lambda_1$, each $[\mathbf{p}(t)]_m$ in \eqref{pFor} increases as $t$ increases.
\end{corollary}

\begin{proof}
    We utilize \eqref{theo2.2} to derive
    \begin{eqnarray}  \label{coro2.1}
        \frac{d\mathbf{p}(t)}{dt}=(2t+1)\mathbf{b}_1+\mathbf{b}_2+\sum_{i=1}^\infty t^{i+1}\mathbf{B}^i\mathbf{b}_1\nonumber\\
        +\sum_{i=1}^\infty(i+1) t^{i}\mathbf{B}^i\big[(t+1)\mathbf{b}_1+\mathbf{b}_2\big]>\mathbf{0}
    \end{eqnarray}
    due to $t>0$, $\mathbf{B}\ge\mathbf{0}$, and $\mathbf{b}_i>\mathbf{0}$. The proof is completed.
\end{proof}

\begin{proposition}  \label{pro3}
    When $0<t<\frac{1}{\lambda_1}$, $[\mathbf{p}(t)]_m\in(0,\infty),m\in\mathcal{M}$ and $\mathbf{1}^T\mathbf{p}(t)\in(0,\infty)$ in \eqref{pFor}.
\end{proposition}

\begin{proof}
    When $t\to0$, $\mathbf{A}(t)=\frac{1}{t}(\mathbf{I}_M-t\mathbf{B})\to\mathbf{I}_M/t$ and $\mathbf{A}^{-1}(t)\to t\mathbf{I}_M$. Then, $\mathbf{p}(t)\to t(t+1)\mathbf{b}_1+t\mathbf{b}_2\to\mathbf{0}$ and $\mathbf{1}^T\mathbf{p}(t)\to0$.

    When $t\to\frac{1}{\lambda_1}$, we utilize \eqref{pFor}
    to construct
    \begin{eqnarray}
        \mathbf{p}(t)&=&\overbrace{\frac{1}{\frac{1}{t}-\lambda_1}}^{\to\infty}\sum_{i=1}^K\overbrace{\mathbf{q}_i\mathbf{g}_i^T\big[(t+1)\mathbf{b}_1+\mathbf{b}_2\big]}^{\mathrm{not~tend~to~zero}}+\nonumber\\
        &&\sum_{i=K+1}^M\underbrace{\frac{1}{\frac{1}{t}-\lambda_i}}_{\to\mathrm{a~constant}}\underbrace{\mathbf{q}_i\mathbf{g}_i^T\big[(t+1)\mathbf{b}_1+\mathbf{b}_2\big]}_{\mathrm{absolute ~value}<\mathrm{a~constant
        }}\nonumber\\
        &\to&\boldsymbol{\infty},
    \end{eqnarray}
    which means $[\mathbf{p}(t)]_m\to\infty,m\in\mathcal{M}$ due to $\mathbf{p}(t)>\mathbf{0}$ from \textbf{Theorem \ref{theo2}}. Based on \textbf{Corollary \ref{coro2}}, we obtain $[\mathbf{p}(t)]_m\in(0,\infty),m\in\mathcal{M}$ and $\mathbf{1}^T\mathbf{p}(t)\in(0,\infty)$. The proof is completed.
\end{proof}

\begin{remark}   \label{rem2}
    \textbf{Corollary \ref{coro2}} and \textbf{Proposition \ref{pro3}} indicate that as $t$ increases within $(0,\frac{1}{\lambda_1})$, each $[\mathbf{p}(t)]_m$ increases within $(0,\infty)$. Thus, when $\mathbf{b}_2\ge\mathbf{b}_1$ is not satisfied in the constraint \eqref{maxmin5.4}, increasing $t$ can guarantee this constraint.

    In order to increase $t^*$ or $\mathbf{p}^*=\mathbf{p}(t^*)$, we can increase $P_{\mathrm{max}}$ due to $P_{\mathrm{max}}=\mathbf{1}^T\mathbf{p}^*$. Moreover, we may need to increase the upper bound of $t$, i.e., $\frac{1}{\lambda_1}$ or $\frac{1}{\mathbf{\rho}(\mathbf{B})}$, which can be achieved by reducing $\Vert\mathbf{B}\Vert$\footnote{Reducing $\Vert\mathbf{B}\Vert$ means reducing the $\rho^2_{n,2,m}/\rho^2_{n,2,n}$'s in $\mathbf{B}$, which requires that both users in each cluster have a large channel correlation.}, a matrix norm of $\mathbf{B}$, due to $\rho(\mathbf{B})\le\Vert\mathbf{B}\Vert$.
\end{remark}

Based on \textbf{Corollary \ref{coro2}} and \textbf{Proposition \ref{pro3}}, we can know that if there is a global maximizer for problem \eqref{maxmin5}, the global maximizer is unique. Thus, we now adopt the binary search algorithm in \textbf{Algorithm \ref{alg:A}} to solve problem \eqref{maxmin5}.

\begin{algorithm}
\caption{Binary Search Algorithm for Problem \eqref{maxmin5}} \label{alg:A}
\begin{algorithmic}[1]
\STATE{Initialize $t_{\mathrm{min}}=0$ and $t_{\mathrm{max}}=\frac{1}{\lambda_1}$.}
\WHILE{$t_{\mathrm{max}}-t_{\mathrm{min}}>\epsilon$}
   \STATE{Let $t=(t_{\mathrm{min}}+t_{\mathrm{max}})/2$ and calculate $\mathbf{p}(t)$ using \eqref{pFor}.}
   \IF{$\mathbf{1}^{T}\mathbf{p}(t)>P_{\mathrm{max}}$}
   \STATE{$t_{\mathrm{max}}=t$.}
   \ELSE
   \STATE{$t_{\mathrm{min}}=t$.}
    \ENDIF
\ENDWHILE
\STATE{Calculate $\mathbf{p}^*=\mathbf{p}\big((t_{\mathrm{min}}+t_{\mathrm{max}})/2\big)$ using \eqref{pFor}.}
\IF{$\mathbf{B}\mathbf{p}^*+\mathbf{b}_2\ge\mathbf{b}_1$}
    \STATE{Calculate and output $(\mathbf{p}^*,\boldsymbol{\Theta}^*)$ using \eqref{the01.1}.}
\ELSE
    \STATE{Output $\mathrm{``no~solution~to~the~problem"}$.}
\ENDIF
\end{algorithmic}
\end{algorithm}

\section{Min-Max Fairness for Partial CSIT}

In this section, we aim to solve problem \eqref{minmax1}. To begin with, we transform $\mathrm{SINR}_{n,i}^2\ge\gamma$ and $\mathrm{SINR}_{n,1}\ge\gamma$ in \eqref{oup} into
\begin{subequations}
\begin{eqnarray}
    \xi_{n,i}\ge\gamma_{n,i}^2=\frac{\frac{1}{\sigma_{n,i}^2\rho^2_{n,i,n}}}{\frac{p_{n,2}}{\gamma}-p_{n,1}-\sum_{m\ne n}p_m\frac{\rho^2_{n,i,m}}{\rho^2_{n,i,n}}}>0,\\
    \xi_{n,1}\ge\gamma_{n,1}=\frac{\frac{1}{\sigma^2_{n,1}\rho^2_{n,1,n}}}{\frac{p_{n,1}}{\gamma}-\sum_{m\ne n}p_m\frac{\rho^2_{n,1,m}}{\rho^2_{n,1,n}}}>0,~~~~~~~~
\end{eqnarray}
\end{subequations}
and simplify \eqref{oup} into
\begin{subequations}
    \begin{eqnarray}
        P_{n,1}^{\mathrm{O}}&=&\mathcal{P}\big\{0<\xi_{n,1}<\max\{\gamma_{n,1}^2,\gamma_{n,1}\}\big\},\label{min2.1}\\
        P_{n,2}^{\mathrm{O}}&=&\mathcal{P}\{0<\xi_{n,2}<\gamma_{n,2}^2\}.
    \end{eqnarray}
\end{subequations}
Then, we use the monotonically increasing relation of $\mathcal{P}(0<\xi_{n,i}<u)$ and $u$ to simplify problem \eqref{minmax1} into
\begin{subequations}  \label{minmax2}
    \begin{eqnarray}
\min_{\mathbf{P},\mathbf{W}}\max_{n\in\mathcal{M}}\gamma^{\mathrm{max}}_n(\mathbf{P},\mathbf{W})~~~~~~~~~~~~~\\
    \mathrm{s.t.}~~~~~\mathbf{P}\ge\mathbf{0},\mathbf{1}^T\mathbf{P}\le P_{\mathrm{max}}~~\\
        \Vert\mathbf{w}_n\Vert=1,n\in\mathcal{M}~~~~~
    \end{eqnarray}
\end{subequations}
where $\gamma^{\mathrm{max}}_n(\mathbf{P},\mathbf{W}
)=\max\big\{\gamma_{n,1}^2,\gamma_{n,1},\gamma_{n,2}^2\big\}>0$. 
Notice that manipulating $1/\gamma_{n}^{\mathrm{max}}(\mathbf{P},\mathbf{W})$ is more convenient, so we further simplify problem \eqref{minmax2} into the max-min problem:
\begin{subequations}  \label{minmax3}
    \begin{eqnarray}
\max_{\mathbf{P},\mathbf{W}}\min_{n\in\mathcal{M}}\gamma^{\mathrm{min}}_n(\mathbf{P},\mathbf{W}
)~~~~~~~~~~~~~~\\
    \mathrm{s.t.}~~~~~\mathbf{P}\ge\mathbf{0},\mathbf{1}^T\mathbf{P}\le P_{\mathrm{max}}~~\\
        \Vert\mathbf{w}_n\Vert=1,n\in\mathcal{M}~~~~~
    \end{eqnarray}
\end{subequations}
where $\gamma^{\mathrm{min}}_n(\mathbf{P},\mathbf{W})=\min\big\{\tilde{\gamma}_{n,1}^2,\tilde{\gamma}_{n,1},\tilde{\gamma}_{n,2}^2\big\}>0$ with
\begin{subequations}
\begin{eqnarray}
\tilde{\gamma}_{n,i}^2=\sigma_{n,i}^2\rho^2_{n,i,n}\Big(\frac{p_{n,2}}{\gamma}-p_{n,1}-\sum_{m\ne n}p_m\frac{\rho^2_{n,i,m}}{\rho^2_{n,i,n}}\Big)>0,\nonumber\\
    \tilde{\gamma}_{n,1}=\sigma^2_{n,1}\rho^2_{n,1,n}\Big(\frac{p_{n,1}}{\gamma}-\sum_{m\ne n}p_m\frac{\rho^2_{n,1,m}}{\rho^2_{n,1,n}}\Big)>0.~~~~~~~\nonumber
\end{eqnarray}   
\end{subequations}

Likewise, this problem is also complicated due to the coupled nature. For simplifying this problem, we first design the precoding matrix to obtain a power control problem.

\subsection{Precoding Matrix Design}

Since both $\tilde{\gamma}_{n,1}^2$ and $\tilde{\gamma}_{n,1}$ in $\gamma_n^{\mathrm{min}} (\mathbf{P},\mathbf{W})$ are associated with $\mathrm{UE}_{n,1}$, we  design $\mathbf{W}$ to eliminate or mitigate the inter-cluster interference of each $\mathrm{UE}_{n,1}$. Similarly, we adopt the ZFBF based on the available strong users' CDI
\begin{eqnarray}
    \mathbf{D}_1=\begin{bmatrix}
        \mathbf{d}_{1,1}^H \\
        \vdots \\
        \mathbf{d}_{M,1}^H
    \end{bmatrix}\in\mathbb{C}^{M\times N}
\end{eqnarray}
and design the precoding matrix as
\begin{eqnarray}
    \mathbf{W}=\mathbf{D}_1^H(\mathbf{D}_1\mathbf{D}_1^H)^{-1}\cdot\mathrm{diag}\Big\{\frac{1}{\Vert\mathbf{a}_1\Vert},\dots,\frac{1}{\Vert\mathbf{a}_M\Vert}\Big\},
\end{eqnarray}
where $\mathbf{a}_m$ is the $m$th column of $\mathbf{D}_1^H(\mathbf{D}_1\mathbf{D}_1^H)^{-1}$ and $\mathbf{w}_m=\mathbf{a}_m/\Vert\mathbf{a}_M\Vert$, and
\begin{eqnarray}  \label{zfbf2}
\vert\mathbf{d}_{i,1}^H\mathbf{w}_j\vert^2=\rho^2_{i,1,j}=\left\{
\begin{array}{ll}
0, & i\ne j, \\
1/\Vert\mathbf{a}_i\Vert^2, & i=j.
\end{array}
\right.
\end{eqnarray}

Utilizing \eqref{zfbf2} and the power allocation parameter $\boldsymbol{\beta}_n$, we can accordingly reduce $\gamma^{\mathrm{min}}_n(\mathbf{P},\mathbf{W})$ to $\gamma^{\mathrm{min}}_n(\mathbf{p},\boldsymbol{\beta}_n)=\min\big\{\tilde{\gamma}_{n,1}^2(p_n,\boldsymbol{\beta}_n),\tilde{\gamma}_{n,1}(p_n,\boldsymbol{\beta}_n),\tilde{\gamma}_{n,2}^2(\mathbf{p},\mathbf{\beta}_n)\big\}$ with
\begin{subequations}
\begin{eqnarray}
\tilde{\gamma}_{n,1}^2(p_n,\boldsymbol{\beta}_n)=\sigma_{n,1}^2\rho^2_{n,1,n}p_n\Big(\frac{\beta_{n,2}}{\gamma}-\beta_{n,1}\Big)>0,~~~~~~~~~~~~~~~~\nonumber\\
    \tilde{\gamma}_{n,1}(p_n,\boldsymbol{\beta}_n)=\sigma^2_{n,1}\rho^2_{n,1,n}\frac{p_n\beta_{n,1}}{\gamma}>0\nonumber,~~~~~~~~~~~~~~~~~~~~~~~~~~~~~\\
    \tilde{\gamma}_{n,2}^2(\mathbf{p},\boldsymbol{\beta}_n)=\sigma_{n,2}^2\rho^2_{n,2,n}\Big(\frac{p_n\beta_{n,2}}{\gamma}-p_n\beta_{n,1}-\sum_{m\ne n}p_m\frac{\rho^2_{n,2,m}}{\rho^2_{n,2,n}}\Big).\nonumber
\end{eqnarray}   
\end{subequations}
Here, because $\mathrm{UE}_{n,1}$ is the strong user due to $d_{n,1}\le d_{n,2}$, i.e., $\sigma^2_{n,1}\ge\sigma^2_{n,2}>0$, $\sigma^2_{n,1}\rho^2_{n,1,n}\ge\sigma^2_{n,2}\rho^2_{n,2,n}>0$ and
\begin{subequations}
\begin{eqnarray}  \label{cons2}
    &\tilde{\gamma}_{n,1}^2(p_n,\boldsymbol{\beta}_n)\ge\tilde{\gamma}_{n,2}^2(\mathbf{p},\boldsymbol{\beta}_n)>0,n\in\mathcal{M}\\
    &\Leftrightarrow(\mathbf{J}+\mathbf{B})\mathbf{p}\ge\mathbf{0},
\end{eqnarray}
\end{subequations}
where $a_n=(1-\sigma^2_{n,2}\rho^2_{n,2,n}/\sigma^2_{n,1}\rho^2_{n,1,n})(\beta_{n,2}/\gamma-\beta_{n,1})$ and $\mathbf{J}=\mathrm{diag}\{a_1,\dots,a_M\}$, always hold in practice. This helps to simplify problem \eqref{minmax3} into the power control problem:
\begin{subequations}  \label{minmax4}
    \begin{eqnarray}
F(P_{\mathrm{max}})=\max_{\mathbf{p},\boldsymbol{\Theta}}\min_{n\in\mathcal{M}}\gamma^{\mathrm{min}}_n(\mathbf{p},\boldsymbol{\beta}_n)~~~~~~~\\
    \mathrm{s.t.}~~~~\mathbf{p}\ge\mathbf{0},\mathbf{1}^T\mathbf{p}\le P_{\mathrm{max}}~~~~~~~~~\\
    \boldsymbol{\beta}_n\ge\mathbf{0},\mathbf{1}^T\boldsymbol{\beta}_n=1,n\in\mathcal{M}\\
    (\mathbf{J}+\mathbf{B})\mathbf{p}\ge\mathbf{0}~~~~~~~~~~~~~~~~
    \end{eqnarray}
\end{subequations}
where $\gamma^{\mathrm{min}}_n(\mathbf{p},\boldsymbol{\beta}_n)=\min\big\{\tilde{\gamma}_{n,1}(p_n,\boldsymbol{\beta}_n),\tilde{\gamma}_{n,2}^2(\mathbf{p},\boldsymbol{\beta}_n)\big\}$.

\subsection{Power Control}

In order to simplify problem \eqref{minmax4}, we provide the following lemmas.

\begin{lemma}    \label{lem6}
    If $(\mathbf{p}^*,\boldsymbol{\Theta}^*)$ is a global maximizer of problem \eqref{minmax4}, we then have
    \begin{eqnarray}  \label{lem6.1}
        p_n^*\ne0,\beta_{n,i}^*\ne0,\forall n\in\mathcal{M},i\in\mathcal{I}.
    \end{eqnarray}
\end{lemma}

\begin{proof}
    This proof is simple because when the constraint \eqref{lem6.1} does not hold, some user's SINR is identically zero and the outage probability is one. The proof is completed.
\end{proof}

\begin{lemma}  \label{lem7}
    If $(\mathbf{p}^*,\boldsymbol{\Theta}^*)$ is a global maximizer of problem \eqref{minmax4}, we then have $\mathbf{1}^T\mathbf{p}^*=P_{\mathrm{max}}$.
\end{lemma}

\begin{proof}
This proof is similar to that of \textbf{Lemma \ref{lem1}}.
\end{proof}

\begin{lemma}  \label{lem8}
    If $(\mathbf{p}^*,\boldsymbol{\Theta}^*)$ is a global maximizer of problem \eqref{minmax4}, we then have
    \begin{eqnarray}  \label{lem8.1}
        F(P_{\mathrm{max}})=\gamma_n^{\mathrm{min}}(\mathbf{p}^*,\boldsymbol{\beta}_n^*),\forall n\in\mathcal{M}.
    \end{eqnarray}
\end{lemma}

\begin{proof}
    We use the proof by contradiction. If the equality \eqref{lem8.1} does not hold, after sorting we can assume
    \begin{eqnarray}  \label{lem8.2}
        F(P_{\mathrm{max}})=\gamma_1^{\mathrm{min}}(\mathbf{p}^*,\boldsymbol{\beta}_1^*)=\cdots=\gamma_i^{\mathrm{min}}(\mathbf{p}^*,\boldsymbol{\beta}_i^*)\nonumber\\<\gamma_{i+1}^{\mathrm{min}}(\mathbf{p}^*,\boldsymbol{\beta}_{i+1}^*)\le\cdots\le\gamma_M^{\mathrm{min}}(\mathbf{p}^*,\boldsymbol{\beta}_M^*).
    \end{eqnarray}
    Then, we set $\sum_{m=1}^ip_m^*=P_f>0$, $\sum_{m=i+1}^M=P_s>0$, and $\mu P_f+\eta P_s=\sum_{m=1}^Mp_m^*$ where $\mu\in\big[1,(\sum_{m=1}^Mp_m^*)/P_f\big)$ and $\eta=(\sum_{m=1}^Mp_m^*-\mu P_f)/P_s\in(0,1]$ decreases as $\mu$ increases. On this foundation, let $\mathbf{p}(\mu)=[\mu p_1^*~\cdots~\mu p_i^*~\eta p_{i+1}^*~\cdots~\eta p_M^*]^T$ and as $\mu$ increases,
    \begin{eqnarray}
        \gamma_n^{\mathrm{min}}\big(\mathbf{p}(\mu),\boldsymbol{\beta}_n^*\big)=\min\Big\{\sigma^2_{n,1}\rho^2_{n,1,n}(\mu p_n^*)\beta_{n,1}^*/\gamma,~~~~~~~~\nonumber\\
        \sigma^2_{n,2}\rho^2_{n,2,n}\Big[-\sum_{m=i+1}^M(\eta p_m^*)\frac{\rho^2_{n,2,m}}{\rho^2_{n,2,n}}+~~~~~~\nonumber\\
        \mu\Big(\underbrace{\frac{p_n^*\beta_{n,2}^*}{\gamma}- p_n^*\beta_{n,1}^*-\sum_{m=1,m\ne n}^i p_m^*\frac{\rho^2_{n,2,m}}{\rho^2_{n,2,n}}}_{\ge0}\Big)\Big]\Big\},n\le i
    \end{eqnarray}
    increases but
    \begin{eqnarray}
        \gamma_n^{\mathrm{min}}\big(\mathbf{p}(\mu),\boldsymbol{\beta}_n^*\big)=\min\Big\{\sigma^2_{n,1}\rho^2_{n,1,n}(\eta p_n^*)\beta_{n,1}^*/\gamma,~~~~~~~~~~~\nonumber\\
        \sigma^2_{n,2}\rho^2_{n,2,n}\Big[-\sum_{m=1}^i(\mu p_m^*)\frac{\rho^2_{n,2,m}}{\rho^2_{n,2,n}}+~~~~~~~~~~~\nonumber\\
        \eta\Big(\underbrace{\frac{p_n^*\beta_{n,2}^*}{\gamma}- p_n^*\beta_{n,1}^*-\sum_{m=i+1,m\ne n}^M p_m^*\frac{\rho^2_{n,2,m}}{\rho^2_{n,2,n}}}_{\ge0}\Big)\Big]\Big\},n>i~~
    \end{eqnarray}
    decreases. Hence, there must exist $\mu_0\in\big(1,(\sum_{m=1}^Mp_m^*)/P_f\big)$ and $\eta_0=(\sum_{m=1}^Mp_m^*-\mu_0P_f)/P_s$ such that
    \begin{eqnarray}
        F(P_{\mathrm{max}})=\gamma_l^{\mathrm{min}}\big(\mathbf{p}(1),\boldsymbol{\beta}_l^*\big)<\gamma_l^{\mathrm{min}}\big(\mathbf{p}(\mu_0),\boldsymbol{\beta}_l^*\big)\nonumber\\
        \le\gamma_k^{\mathrm{min}}\big(\mathbf{p}(\mu_0),\boldsymbol{\beta}_k^*\big),\forall l\le i,k>i~~~~~~
    \end{eqnarray}
    which means that $\min_{n\in\mathcal{M}}\gamma_n^{\mathrm{min}}\big(\mathbf{p}(\mu_0),\boldsymbol{\beta}_n^*\big)>F(P_{\mathrm{max}})$ and $(\mathbf{p}^*,\boldsymbol{\Theta}^*)$ is not a global maximizer of problem \eqref{minmax4}. This contradicts the assumption and the proof is completed.
\end{proof}

Utilizing \textbf{Lemma \ref{lem6}},  \textbf{Lemma \ref{lem7}}, and  \textbf{Lemma \ref{lem8}}, we simplify problem \eqref{minmax4} into
\begin{subequations}  \label{minmax5}
    \begin{eqnarray}
F(P_{\mathrm{max}})=\max_{\mathbf{p},\boldsymbol{\Theta},y}y~~~~~~~~~~~~~~~~~~~~~~~~~~\\
    \mathrm{s.t.}~~~~~\gamma^{\mathrm{min}}_n(\mathbf{p},\boldsymbol{\beta}_n)=y,n\in\mathcal{M}~~~~\\
    \mathbf{p}>\mathbf{0},\mathbf{1}^T\mathbf{p}=P_{\mathrm{max}}~~~~~~~~~\\
    \boldsymbol{\beta}_n>\mathbf{0},\mathbf{1}^T\boldsymbol{\beta}_n=1,n\in\mathcal{M}\\
    (\mathbf{J}+\mathbf{B})\mathbf{p}\ge\mathbf{0}.~~~~~~~~~~~~~~~
    \end{eqnarray}
\end{subequations}
To further simplify problem \eqref{minmax5}, we provide \textbf{Lemma \ref{lem9}}.

\begin{lemma}  \label{lem9}
    If $(\mathbf{p}^*,\boldsymbol{\Theta}^*,y^*)$ is a global maximizer of problem \eqref{minmax5}, we then have
    \begin{eqnarray}  \label{lem9.1}
        F(P_{\mathrm{max}})=y^*=\tilde{\gamma}_{n,1}(p_n^*,\boldsymbol{\beta}_n^*)=\tilde{\gamma}_{n,2}^2(\mathbf{p}^*,\boldsymbol{\beta}_n^*),\forall n\in\mathcal{M}.
    \end{eqnarray}
\end{lemma}

\begin{proof}
    We still use the proof by contradiction. If the equality \eqref{lem9.1} does not hold, there must exist $k\in\mathcal{M}$ such that either $\tilde{\gamma}_{k,1}(p_k^*,\boldsymbol{\beta}_k^*)>\tilde{\gamma}_{k,2}^2(\mathbf{p}^*,\boldsymbol{\beta}_k^*)$ or $\tilde{\gamma}_{k,1}(p_k^*,\boldsymbol{\beta}_k^*)<\tilde{\gamma}_{k,2}^2(\mathbf{p}^*,\boldsymbol{\beta}_k^*)$. Then, note that as $\beta_{k,1}$ increases, $\tilde{\gamma}_{k,1}(p_k^*,\boldsymbol{\beta}_k)=\sigma^2_{k,1}\rho^2_{k,1,k}p_k^*\beta_{k,1}/\gamma$ increases but
    \begin{eqnarray}
        \tilde{\gamma}_{k,2}^2(\mathbf{p}^*,\boldsymbol{\beta}_k)=\sigma^2_{k,2}\rho^2_{k,2,k}\Big(\frac{p_k^*(1-\beta_{k,1})}{\gamma}-p_k^*\beta_{k,1}\nonumber\\
        -\sum_{m\ne k}p_m^*\frac{\rho^2_{k,2,m}}{\rho^2_{k,2,k}}\Big)
    \end{eqnarray}
    decreases. Moreover, \textbf{Lemma \ref{lem6}} indicates $\beta_k^*\in(0,1)$. Hence, when $\tilde{\gamma}_{k,1}(p_k^*,\boldsymbol{\beta}_k^*)>\tilde{\gamma}_{k,2}^2(\mathbf{p}^*,\boldsymbol{\beta}_k^*)$, there exists $\boldsymbol{\beta}_k^0=[\beta_{k,1}^0~\beta_{k,2}^0]^T$ with $\beta_{k,1}^0\in(0,\beta_{k,1}^*)$ such that
    \begin{eqnarray}
        \tilde{\gamma}_{k,1}(p_k^*,\boldsymbol{\beta}_k^*)>(p_k^*,\boldsymbol{\beta}_k^0)
        \ge\underbrace{\tilde{\gamma}_{k,2}^2(\mathbf{p}^*,\boldsymbol{\beta}_k^0)}_{=\gamma_k^{\mathrm{min}}(\mathbf{p}^*,\boldsymbol{\beta}_k^0)}>\underbrace{\tilde{\gamma}_{k,2}^2(\mathbf{p}^*,\boldsymbol{\beta}_k^*)}_{=\gamma_k^{\mathrm{min}}(\mathbf{p}^*,\boldsymbol{\beta}_k^*)}
    \end{eqnarray}
    which means that $\gamma_k^{\mathrm{min}}(\mathbf{p}^*,\boldsymbol{\beta}_k^0)>\gamma_k^{\mathrm{min}}(\mathbf{p}^*,\boldsymbol{\beta}_k^*)=F(P_{\mathrm{max}})$ and $(\mathbf{p}^*,\boldsymbol{\Theta}^*,y^*)$ is not a global maximizer of problem \eqref{minmax5}. This contracts the assumption.

    When $\tilde{\gamma}_{k,1}(p_k^*,\boldsymbol{\beta}_k^*)<\tilde{\gamma}_{k,2}^2(\mathbf{p}^*,\boldsymbol{\beta}_k^*)$, we can obtain the same contradiction. The proof is completed.
\end{proof}

\textbf{Lemma \ref{lem8}} and \textbf{Lemma \ref{lem9}} also verify that max-min fairness
always causes absolute user fairness \cite{Timotheou}, i.e., all users’ outage probabilities are the same $\mathcal{P}\{0<\xi_{n,i}<1/y^*\}$. From \textbf{Lemma \ref{lem9}}, we obtain \textbf{Theorem \ref{theo3}} and \textbf{Corollary \ref{coro3}}.

\begin{theorem}  \label{theo3}
    If $(\mathbf{p}^*,\boldsymbol{\Theta}^*,y^*)$ is a global maximizer of problem \eqref{minmax5}, we then have
    \begin{eqnarray}  \label{theo3.1}
        \beta_{n,1}^*=\frac{1-\gamma\sum_{m\ne n}\frac{p_m^*}{p_n^*}\frac{\rho^2_{n,2,m}}{\rho^2_{n,2,n}}}{\frac{\sigma^2_{n,1}\rho^2_{n,1,n}}{\sigma^2_{n,2}\rho^2_{n,2,n}}+1+\gamma}>0,\forall n\in\mathcal{M}.
    \end{eqnarray}
\end{theorem}

\begin{proof}
    Let $\tilde{\gamma}_{n,1}(p_n^*,\boldsymbol{\beta}_n)=\tilde{\gamma}_{n,2}^2(p_n^*,\boldsymbol{\beta}_n)$ and obtain a linear equation with respect to $\beta_{n,1}$. Solving this equation obtains \eqref{theo3.1} and the proof is completed.
\end{proof}

\begin{corollary}    \label{coro3}
    If $(\mathbf{p}^*,\boldsymbol{\Theta}^*,y^*)$ is a global maximizer of problem \eqref{minmax5} and $\gamma>(1-\frac{\sigma^2_{n,1}\rho^2_{n,1,n}}{\sigma^2_{n,2}\rho^2_{n,2,n}})/(2\sum_{m\ne n}\frac{p_m^*}{p_n^*}\frac{\rho^2_{n,2,m}}{\rho^2_{n,2,n}}+1)$, we then have $\beta_{n,1}^*<\frac{1}{2}$.
\end{corollary}

\begin{proof}
    $\beta_{n,1}^*<\frac{1}{2}$ in \eqref{theo3.1} results in this result. The proof is completed.
\end{proof}

\textbf{Corollary \ref{coro3}} indicates that in practice, $\frac{\sigma^2_{n,1}\rho^2_{n,1,n}}{\sigma^2_{n,2}\rho^2_{n,2,n}}\ge1,\forall n\in\mathcal{M}$ always holds, so $\beta_{n,1}^*<\frac{1}{2},\forall n\in\mathcal{M}$ always holds and max-min fairness can guarantee the implicit constraint in NOMA. Utilizing \textbf{Theorem \ref{theo3}}, we simplify problem \eqref{minmax5} into
\begin{subequations}  \label{minmax6}
    \begin{eqnarray}
F(P_{\mathrm{max}})=\max_{\mathbf{p},y}y~~~~~~~~~~~~~~~~~~~~~~~~~~\\
    \mathrm{s.t.}~~~~~\gamma^{\mathrm{min}}_n(\mathbf{p})=y,n\in\mathcal{M}\label{minmax6.2}~~~~~~~~\\
    \mathbf{p}>\mathbf{0},\mathbf{1}^T\mathbf{p}=P_{\mathrm{max}}~~~~~~~~~\\
    \boldsymbol{\beta}_n>\mathbf{0},\mathbf{1}^T\boldsymbol{\beta}_n=1,n\in\mathcal{M}\\
    (\mathbf{J}+\mathbf{B})\mathbf{p}\ge\mathbf{0}.~~~~~~~~~~~~~~~\label{minmax6.4}
    \end{eqnarray}
\end{subequations}
where the feasible region of $y$ is $(0,\infty)$ and
\begin{eqnarray}
    \gamma^{\mathrm{min}}_n(\mathbf{p})=\sigma^2_{n,1}\rho^2_{n,1,n}\frac{\frac{p_n}{\gamma}-\sum_{m\ne n}p_m\frac{\rho^2_{n,2,m}}{\rho^2_{n,2,n}}}{\frac{\sigma^2_{n,1}\rho^2_{n,1,n}}{\sigma^2_{n,2}\rho^2_{n,2,n}}+1+\gamma}>0.\nonumber
\end{eqnarray}

\subsection{Closed-Form or Iterative Solution}

To solve problem \eqref{minmax6}, we first focus more on the equality constraint \eqref{minmax6.2} and transform it into
\begin{eqnarray}
    \frac{p_n}{\gamma}-\sum_{m\ne n}p_m\frac{\rho^2_{n,2,m}}{\rho^2_{n,2,n}}=\frac{y(1+\gamma)}{\sigma^2_{n,1}\rho^2_{n,1,n}}+\frac{y}{\sigma^2_{n,2}\rho^2_{n,2,n}},n\in\mathcal{M}
\end{eqnarray}
and the compact matrix form
\begin{eqnarray}  \label{sle}
    \mathbf{G}(\gamma)\mathbf{p}=y(1+\gamma)\mathbf{k}_1+y\mathbf{k}_2>\mathbf{0}
\end{eqnarray}
where $\mathbf{G}(\gamma)=\frac{\mathbf{I}_M}{\gamma}-\mathbf{B}$, $\mathbf{k}_i=\big[\frac{1}{\sigma^2_{1,i}\rho^2_{1,i,1}}~\cdots~\frac{1}{\sigma^2_{M,i}\rho^2_{M,i,M}}\big]^T>\mathbf{0},i\in\mathcal{I}$.

Because the system of linear equations \eqref{sle} is similar to \eqref{SLs},
we give the following theorem, corollary and proposition about the solution:
\begin{eqnarray}  \label{sle2}
    \mathbf{p}(\gamma,y)=y\mathbf{G}^{-1}(\gamma)\big[(1+\gamma)\mathbf{k}_1+\mathbf{k}_2\big]~~~~~~~~~~~~~\nonumber\\
    =y(1+\gamma)\sum_{i=1}^M\frac{\mathbf{q}_i\mathbf{g}_i^T\mathbf{k}_1}{\frac{1}{\gamma}-\lambda_i}+y\sum_{i=1}^M\frac{\mathbf{q}_i\mathbf{g}_i^T\mathbf{k}_2}{\frac{1}{\gamma}-\lambda_i},y>0
\end{eqnarray}
where $\mathbf{G}^{-1}(\gamma)=\mathbf{Q}\mathrm{diag}\big\{(\frac{1}{\gamma}-\lambda_1)^{-1},\dots,(\frac{1}{\gamma}-\lambda_M)^{-1}\big\}\mathbf{Q}^{-1}$.

\begin{theorem}  \label{theo4}
    $\mathbf{p}(\gamma,y)>\mathbf{0}$ in \eqref{sle2} if and only if $\rho(\mathbf{B})<\frac{1}{\gamma}$ or $0<\gamma<\frac{1}{\lambda_1}$.
\end{theorem}

\begin{proof}
    This proof is similar to that of \textbf{Theorem \ref{theo2}}.
\end{proof}

\begin{corollary}    \label{coro4}
    When $0<\gamma<1/\lambda_1$ and $y\in(0,\infty)$, each $[\mathbf{p}(\gamma,y)]_m$ in \eqref{sle2} increases as $\gamma$ or $y$ increases.
\end{corollary}

\begin{proof}
    This proof is similar to that of \textbf{Corollary \ref{coro2}}.
\end{proof}

\begin{proposition}     \label{pro4}
    When $0<\gamma<1/\lambda_1$ and $y\in(0,\infty)$, $[\mathbf{p}(\gamma,y)]_m\in(0,\infty),m\in\mathcal{M}$ and $\mathbf{1}^T\mathbf{p}(\gamma,y)\in(0,\infty)$ in \eqref{sle2}.
\end{proposition}

\begin{proof}
    This proof is similar to that of \textbf{Proposition \ref{pro3}}.
\end{proof}

\begin{remark}  \label{rem3}
    After user clustering and the ZFBF based on the strong users' CSI $\mathbf{H}_1$ or CDI $\mathbf{D}_1$, $\mathbf{B}$ is determined and \textbf{Remark \ref{rem2}} indicates that $t$, the same SINR for all users, has an upper bound $\frac{1}{\lambda_1}$ without the power constraint\footnote{The SINR of $\mathrm{UE}_{n,i}$ is a function of $p_n\Gamma_{n,i}$ and under partial CSIT, the BS does not know the CQI $\Gamma_{n,i}$, so $p_n\Gamma_{n,i}$ is a random variable and can take any value in $(0,\infty)$ in the BS side, which is equivalent to that $p_n$ can take any value in $(0,\infty)$ with fixed $\Gamma_{n,i}$ under perfect CSIT.}. This is consistent with that $\gamma$, the same target SINR for all users, has the same upper bound $\frac{1}{\lambda_1}$ from \textbf{Theorem \ref{theo4}}, which guarantees the same outage probability $\mathcal{P}\{0<\xi_{n,i}<1/y^*\}<1,\forall n\in\mathcal{M},i\in\mathcal{I}$ for all users.

    \textbf{Corollary \ref{coro4}} and \textbf{Proposition \ref{pro4}} indicate that as $\gamma$ increases within $(0,\frac{1}{\lambda_1})$ or $y$ increases within $(0,\infty)$, each $[\mathbf{p}(\gamma,y)]_m$ increases within $(0,\infty)$. Thus, given $\mathbf{1}^T\mathbf{p}(\gamma,y^*)=P_{\mathrm{max}}$, $y^*$ decreases (i.e. $\mathcal{P}\{0<\xi_{n,i}<1/y^*\}$ increases) as $\gamma$ increases.
\end{remark}

Based on \textbf{Corollary \ref{coro4}} and \textbf{Proposition \ref{pro4}}, we can know that given $\gamma\in(0,\frac{1}{\lambda_1})$, if there is a global maximizer for problem \eqref{minmax6}, the global maximizer is unique. Thus, we can utilize \eqref{sle2} and $\mathbf{1}^T\mathbf{p}(\gamma,y^*)=P_{\mathrm{max}}$ to derive the closed-form solution:
\begin{subequations}
\begin{eqnarray}
    y^*&=&P_{\mathrm{max}}/\big(\mathbf{1}^T\mathbf{G}^{-1}(\gamma)[(1+\gamma)\mathbf{k}_1+\mathbf{k}_2]\big),\label{y*}\\
    \mathbf{p}^*&=&y^*\mathbf{G}^{-1}(\gamma)[(1+\gamma)\mathbf{k}_1+\mathbf{k}_2].
\end{eqnarray}
\end{subequations}
Here, if $(\mathbf{J}+\mathbf{B})\mathbf{p}^*\ge\mathbf{0}$, calculate and output $(\mathbf{p}^*,\boldsymbol{\Theta}^*)$ using \eqref{theo3.1}, else output $\mathrm{``no~solution~to~the~problem"}$. Importantly, we utilize \textbf{Theorem \ref{theo5}} to show that the system can obtain full transmit diversity.
\begin{theorem}   \label{theo5}
    Under the proposed user fairness, the system can achieve the greatest diversity gain $N$.
\end{theorem}

\begin{proof}
    Given the proposed user fairness and $\gamma\in(0,\frac{1}{\lambda_1})$, all users share the same outage probability $\mathcal{P}\{0<\xi_{n,i}<1/y^*\}$ and $\mathrm{UE}_{n,i}$’s diversity gain $\mathcal{G}_{n,i}$ is calculated by
    \begin{eqnarray}
        \mathcal{G}_{n,i}=\lim_{P_\mathrm{max}\to\infty}\frac{-\ln\mathcal{P}\{0<\xi_{n,i}<1/y^*\}}{\ln P_{\mathrm{max}}}~~~~~~~\nonumber\\
        =\lim_{P_\mathrm{max}\to\infty}\frac{-\ln\int_0^{1/y^*}f_{\xi_{n,i}}(x)dx}{\ln \big(y^*\mathbf{1}^T\mathbf{G}^{-1}(\gamma)[(1+\gamma)\mathbf{k}_1+\mathbf{k}_2]\big)}
    \end{eqnarray}
    where $P_{\mathrm{max}}\to\infty\Leftrightarrow y^*\to\infty$ from \eqref{y*}. Therefore, $\mathrm{UE}_{n,i}$’s diversity gain $\mathcal{G}_{n,i}$ is
    \begin{eqnarray}
        \mathcal{G}_{n,i}&=&\lim_{y^*\to\infty}\frac{-\ln\int_0^{1/y^*}f_{\xi_{n,i}}(x)dx}{\ln \big(y^*\mathbf{1}^T\mathbf{G}^{-1}(\gamma)[(1+\gamma)\mathbf{k}_1+\mathbf{k}_2]\big)}\nonumber\\
        &=&\lim_{y^*\to\infty}\frac{-\big[\ln\int_0^{1/y^*}f_{\xi_{n,i}}(x)dx\big]'}{\big[\ln \big(y^*\mathbf{1}^T\mathbf{G}^{-1}(\gamma)[(1+\gamma)\mathbf{k}_1+\mathbf{k}_2]\big)\big]'}\nonumber\\
        &=&\lim_{y^*\to\infty}\frac{f_{\xi_{n,i}}(\frac{1}{y^*})\times\frac{1}{[y^*]^2}}{\int_0^{1/y^*}f_{\xi_{n,i}}(x)dx\times\frac{1}{y^*}}\nonumber\\
    &=&\lim_{\nu=1/y^*\to0}\frac{\big[f_{\xi_{n,i}}(\nu)\times\nu\big]'}{\big[\int_0^{\nu}f_{\xi_{n,i}}(x)dx\big]'}\nonumber\\
    &=&\lim_{\nu\to0}\frac{\big(\frac{\nu^N}{\Gamma(N)}e^{-\nu}\big)'}{\frac{\nu^{N-1}}{\Gamma(N)}e^{-\nu}}=N,n\in\mathcal{M},i\in\mathcal{I}.
    \end{eqnarray}
    The proof is completed.
\end{proof}

On the other hand, if $y^*>0$, i.e., the same $\mathcal{P}\{0<\xi_{n,i}<1/y^*\}$ for all users, is given, we can also adopt the binary search algorithm in \textbf{Algorithm \ref{alg:B}} to find $\gamma^*$.

\begin{algorithm}
\caption{Binary Search Algorithm for Problem \eqref{minmax6}}\label{alg:B}
\begin{algorithmic}[1]
\STATE{Initialize $\gamma_{\mathrm{min}}=0$ and $\gamma_{\mathrm{max}}=\frac{1}{\lambda_1}$.}
\WHILE{$\gamma_{\mathrm{max}}-\gamma_{\mathrm{min}}>\epsilon$}
   \STATE{Let $\gamma=(\gamma_{\mathrm{min}}+\gamma_{\mathrm{max}})/2$ and calculate $\mathbf{1}^T\mathbf{p}(\gamma,y^*)$ using \eqref{sle2}.}
   \IF{$\mathbf{1}^{T}\mathbf{p}(\gamma,y^*)>P_{\mathrm{max}}$}
   \STATE{$\gamma_{\mathrm{max}}=\gamma$.}
   \ELSE
   \STATE{$\gamma_{\mathrm{min}}=\gamma$.}
    \ENDIF
\ENDWHILE
\STATE{Let $\gamma^*=(\gamma_{\mathrm{min}}+\gamma_{\mathrm{max}})/2$ and calculate $\mathbf{p}^*=\mathbf{p}\big(\gamma^*,y^*\big)$ using \eqref{sle2}.}
\IF{$(\mathbf{J}+\mathbf{B})\mathbf{p}^*\ge\mathbf{0}$}
    \STATE{Calculate and output $(\gamma^*,\mathbf{p}^*,\boldsymbol{\Theta}^*)$ using \eqref{theo3.1}.}
\ELSE
    \STATE{Output $\mathrm{``no~solution~to~the~problem"}$.}
\ENDIF
\end{algorithmic}
\end{algorithm}
  


\section{Simulation Results}

Simulation results are used to prove the proposed absolute user fairness. Unless otherwise specified, let $N=5$, $M=3$, $\alpha=2$, $P_{\mathrm{max}}=10^4$, $\epsilon=10^{-5}$, and all users' distances follow $\mathcal{U}(1,400)$.

\subsection{User Clustering}   \label{sr.1}

It is well-known that for the considered downlink multiuser MIMO-NOMA system \cite{kimy2013non}, due to the ZFBF based on the strong users' CSI or CDI, $\mathrm{UE}_{n,1}$ and $\mathrm{UE}_{n,2}$ in the $n$th cluster should have a large channel correlation $\vert\mathbf{d}^H_{n,1}\mathbf{d}_{n,2}\vert$, which can reduce the inter-cluster interference of each $\mathrm{UE}_{n,2}$ and each $\rho^2_{n,2,m}/\rho^2_{n,2,n}$ in $\mathbf{B}$, i.e., reducing $\Vert\mathbf{B}\Vert$ and $\rho(\mathbf{B})$ from \textbf{Remark \ref{rem2}}. Moreover, $\mathrm{UE}_{n,1}$ and $\mathrm{UE}_{n,2}$ should also have a large gain difference $\zeta_{n,1}-\zeta_{n,2}$, where $\zeta_{n,i}=\Gamma_{n,i}$ for perfect CSIT and $\zeta_{n,i}=\sigma^2_{n,i}$ for partial CSIT, which can guarantee the constraint \eqref{maxmin5.4} in problem \eqref{maxmin5} and \eqref{minmax6.4} in problem \eqref{minmax6}.

According to the above analysis, to propose a computationally efficient user clustering algorithm, we leverage gain and channel correlation to perform a two-tier search:
\begin{itemize}
    \item First sort the users based on the gains and group the users into two sets for the strong and weak users, respectively.
    \item Then, sequentially select two users from the two sets into one cluster based on the maximum channel correlation.
\end{itemize}
The details are given in \textbf{Algorithm \ref{alg:C}}\footnote{It is worth pointing out that before user clustering, we should carefully select $2M$ users from numerous users to improve the system performance \cite{kimy2013non}, but this selection procedure is beyond the scope of this paper. This explains why the system performance of simulation results is poor.}.

\begin{algorithm}
\caption{Two-Tier Search Based User Clustering\label{alg:C}}
\begin{algorithmic}[1]
\STATE{Input $2M$ users' available information $(\mathbf{d}^H_i,\zeta_i)$'s.}
\STATE{Sort the $(\mathbf{d}^H_i,\zeta_i)$'s so that $\zeta_1\ge\cdots\ge\zeta_{2M}$.}
\STATE{Let the $i$th user be $\mathrm{UE}_{M+1-i,1}$ for $i=1,\dots,M$.}
\FOR{$i=1,\dots,M$}
   \STATE{Let $\mathrm{cor}=-\infty$.}
   \FOR{$k=M+i,\dots,2M$}
   \IF{$\vert\mathbf{d}^H_{M+1-i}\mathbf{d}_{k}\vert>\mathrm{cor}$}
\STATE{$\mathrm{cor}=\vert\mathbf{d}^H_{M+1-i}\mathbf{d}_{k}\vert$ and exchange the order of $(\mathbf{d}^H_{M+i},\zeta_{M+i})$ and $(\mathbf{d}^H_{k},\zeta_k)$.}
    \ENDIF
    \ENDFOR
    \STATE{Let the $(M+i)$th user be $\mathrm{UE}_{i,2}$.}
\ENDFOR
\end{algorithmic}
\end{algorithm}

\subsection{Max-Min Fairness for Perfect CSIT}

\begin{figure}
\centerline{\includegraphics[width=22pc]{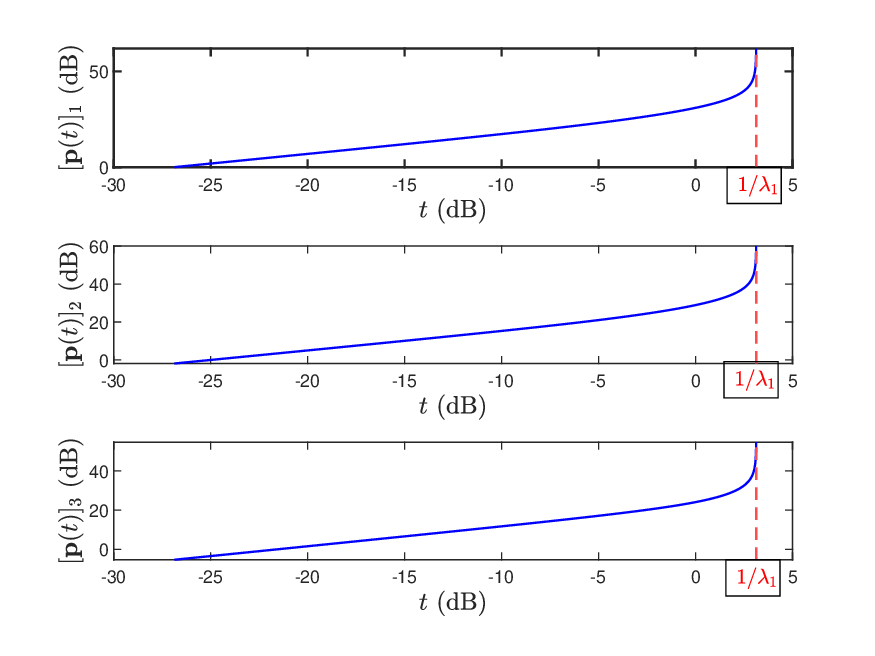}}
\caption{Relation of $t$ and each $[\mathbf{p}(t)]_m$.}\label{f0}
\end{figure}

\begin{figure} 
\centerline{\includegraphics[width=22pc]{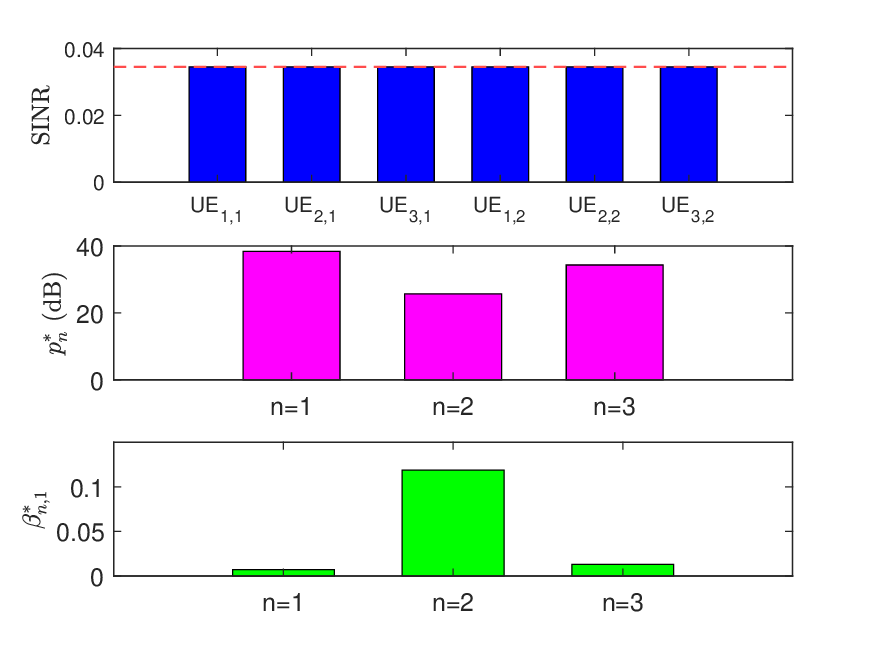}}
\caption{SINR and power allocation comparison of different clusters.}\label{f1}
\end{figure}

Fig. \ref{f0} illustrates the monotonically increasing relation of $t$ and each $[\mathbf{p}(t)]_m$, i.e., as $t$ increases within $(0,1/\lambda_1)$, each $[\mathbf{p}(t)]_m$ increases within $(0, \infty)$, and verifies \textbf{Corollary \ref{coro2}} and \textbf{Proposition \ref{pro3}}. Fig. \ref{f1} compares the SINRs of all users and shows that all users share the same SINR $t^*$, which verifies the proposed absolute user fairness. Moreover, Fig. \ref{f1} shows the $\beta_{n,1}^*$'s of all clusters and verifies that each $\beta_{n,1}^*\in(0,1/2)$ from \textbf{Corollary \ref{coro1}}.

\begin{figure} 
\centerline{\includegraphics[width=20pc]{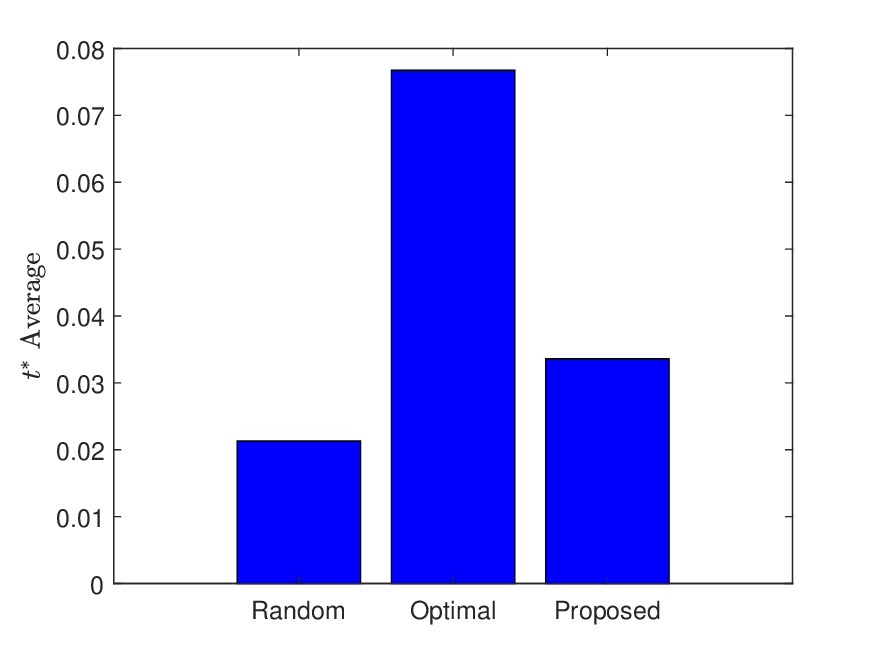}}
\caption{Comparison of three user clustering algorithms.}\label{f2}
\end{figure}

Fig. \ref{f2} compares the random, optimal, and proposed user clustering algorithms by $t^*$ averages. Here, each $t^*$ average is calculated by the same $100$ independent trials. From this figure, we can know that as expected, the optimal user clustering is the best method to maximize $t^*$, the same SINR for all users, but has high complexity of computation when $M$ is large. Moreover, the proposed user clustering with low computational complexity outperforms the random user clustering.

\subsection{Min-Max Fairness for Partial CSIT}

\begin{figure}
\centerline{\includegraphics[width=20pc]{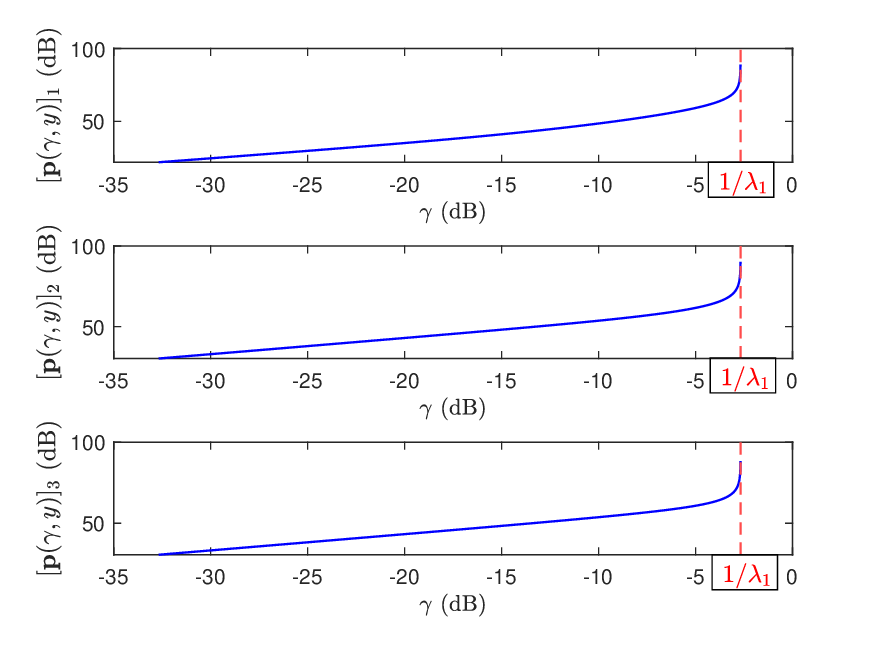}}
\caption{Relation of $\gamma$ and each $[\mathbf{p}(\gamma,y)]_m$ with $y=1$.}\label{f3}
\end{figure}

\begin{figure} 
\centerline{\includegraphics[width=20pc]{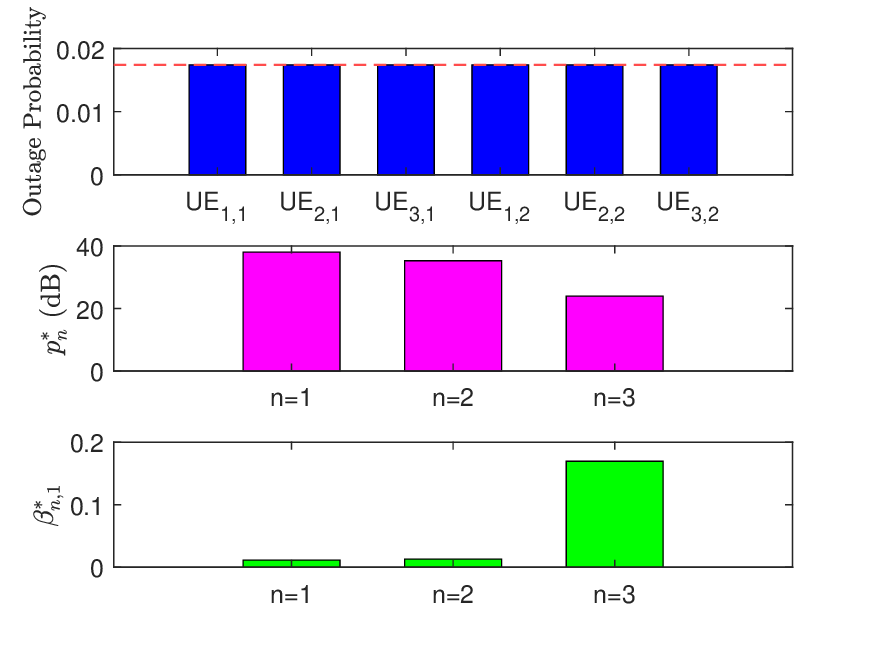}}
\caption{Outage probability and power allocation comparison of different clusters with $\gamma=1/(50\lambda_1)$.}\label{f4}
\end{figure}

Fig. \ref{f3} illustrates the monotonically increasing relation of $\gamma$ and each $[\mathbf{p}(\gamma,y)]_m$, i.e., as $\gamma$ increases within $(0, 1/\lambda_1)$, each $[\mathbf{p}(\gamma,y)]_m$ increases within $(0, \infty)$, and verifies \textbf{Corollary \ref{coro4}} and
\textbf{Proposition \ref{pro4}}. Fig. \ref{f4} compares the outage probabilities of all users and shows that all users share the same outage probability $\mathcal{P}\{0<\xi_{n,i}<1/y^*\}$, which verifies the proposed absolute user fairness. Moreover, Fig. \ref{f4} shows the $\beta_{n,1}^*$'s of all clusters and verifies that each $\beta_{n,1}^*\in(0,1/2)$ from \textbf{Corollary \ref{coro3}}.

\begin{figure} 
\centerline{\includegraphics[width=20pc]{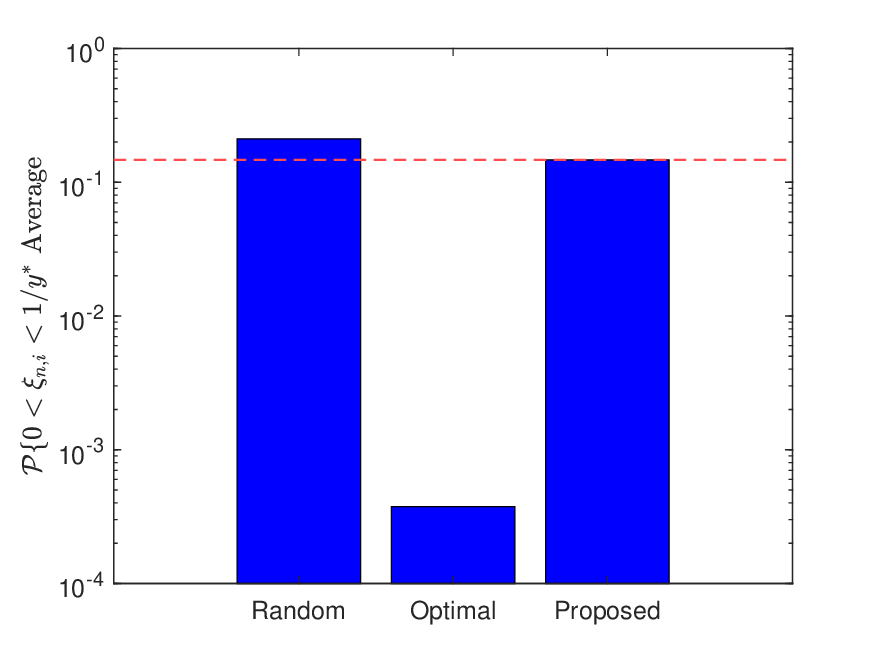}}
\caption{Comparison of three user clustering algorithms with $\gamma=0.005$.}\label{f5}
\end{figure}

Fig. \ref{f5} compares the random, optimal, and proposed user clustering algorithms by $\mathcal{P}\{0<\xi_{n,i}<1/y^*\}$ averages. Here, each $\mathcal{P}\{0<\xi_{n,i}<1/y^*\}$ average is calculated by the same $100$ independent trials. From this figure, we can know that as expected, the optimal user clustering is the best method to minimize $\mathcal{P}\{0<\xi_{n,i}<1/y^*\}$, the same outage probability for all users, but has high complexity of computation when $M$ is large. Moreover, the proposed user clustering with low computational complexity outperforms the random user clustering.

\begin{figure} 
\centerline{\includegraphics[width=20pc]{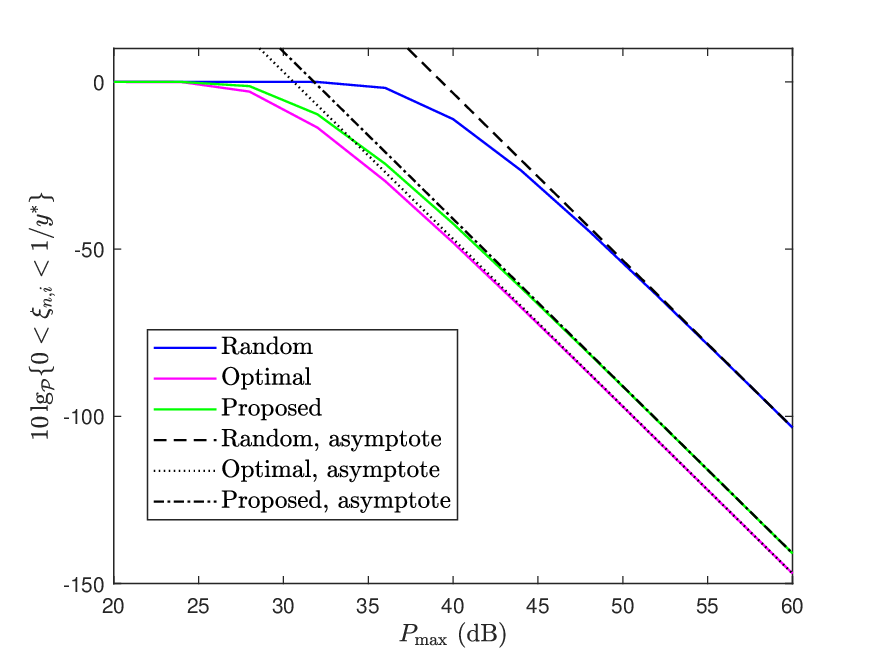}}
\caption{Relation of $10\lg\mathcal{P}\{0<\xi_{n,i}<1/y^*\}$ and $10\lg P_{\mathrm{max}}$ with $\gamma=0.002$.}\label{f6}
\end{figure}

Fig. \ref{f6} illustrates the relation of $10\lg\mathcal{P}\{0<\xi_{n,i}<1/y^*\}$ and $10\lg P_{\mathrm{max}}$. We can see that as expected, under each user clustering, $\mathcal{P}\{0<\xi_{n,i}<1/y^*\}$ decreases as $P_\mathrm{max}$ increases. Importantly, the three asymptotes in this figure share the same  slope $-N$, which indicates that the diversity gain of each user is $N$ and \textbf{Theorem \ref{theo5}} is verified.

\section{CONCLUSION}
Max-min or min-max fairness has been adopted for a downlink multiuser MIMO-NOMA system, where one multi-antenna BS transmits to multiple NOMA clusters and each cluster is composed of two single-antenna users sharing a single beam on the same time-frequency resource block. Specifically, under perfect CSIT, max-min fairness is adopted to maximize the minimum achievable rate among all users. Under partial CSIT, i.e., the BS only knows the distance and CDI of each user, min-max fairness is adopted to minimize the maximum outage probability among all users. Under either CSIT case, to avoid solving the coupled problem, ZFBF has been designed to obtain a power control problem and then a binary search or formula based algorithm has been proposed. Theoretical analyses have revealed that under either CSIT case, the problem can be guaranteed to have a unique solution and the proposed algorithm is computationally efficient due to only simple closed-form expressions involved. Simulation results have verified the proposed absolute user fairness and user clustering algorithm.

\bibliographystyle{IEEEtran}
\bibliography{reference}




\end{document}